\documentclass[12pt]{article}
\usepackage{color}
\usepackage{authblk}
\usepackage{hyperref} 
\usepackage{hanging}

\usepackage{graphicx}
\usepackage{amssymb}
\usepackage{epstopdf}
\usepackage{color}
\usepackage{bm}
\usepackage{hyperref}

\let\ssection=\section
\renewcommand{\section}{\setcounter{equation}{0}\ssection}

\newcommand\mathC{\mkern1mu\raise2.2pt\hbox{$\scriptscriptstyle|$}
        {\mkern-7mu\rm C}}              

\newcommand\be{\begin{equation}}
\newcommand\ee{\end{equation}}

\title{{\bf\large{Peaceful Coexistence: Examining Kent's Relativistic Solution to the Quantum Measurement Problem }}}

\author[1]{Jeremy Butterfield \thanks{Email: jb56@cam.ac.uk}}
\affil[1]{Trinity College, Cambridge, CB2 1TQ, Cambridge, United Kingdom}

\date{}
\begin{document}

\catcode`\'=\active\relax\def'{^{\prime}}\catcode`\'=12\relax

\maketitle

\begin{center}
18 October 2017:  
{\em Dedicated to the memory of Abner Shimony (1928 - 2015)}:\\
Forthcoming in the Proceedings of the 2015 Nagoya Winter Workshop on Reality and Measurement in Quantum Theory 
\end{center}

\begin{abstract}
Can there be `peaceful coexistence' between quantum theory and special relativity? Thirty years ago, Shimony hoped that isolating the culprit (i.e. the false assumption) in proofs of Bell inequalities as Outcome Independence would secure such peaceful coexistence: or, if not secure it, at least show a way---maybe the best or only way---to secure it. 

In this paper, I begin by being sceptical of Shimony's approach, urging that we need a relativistic solution to the quantum measurement problem (Section \ref{preamble2}). Then I analyse  Outcome Independence in Kent's realist one-world Lorentz-invariant interpretation of quantum theory (Section \ref{intro} and \ref{od}). Then I consider Shimony's other condition, Parameter Independence, both in Kent's proposal and more generally, in the light of recent remarkable theorems by Colbeck, Renner and Leegwater (Section \ref{pi?}).

For both Outcome Independence and Parameter Independence, there is a striking analogy with the situation in pilot-wave theory.  Finally, I will suggest that these recent theorems make some kind of peaceful coexistence mandatory for someone who, like Shimony, endorses Parameter Independence.\\ \\


\end{abstract}

\newpage

\tableofcontents

\newpage

\section{Introduction}\label{preamble}
My topic is the assumption of proofs of Bell inequalities that is usually considered `the culprit', i.e. considered to be shown false by the experimental violation of the Bell inequality in question. Thirty years ago, Shimony (1984, 1986) argued that  denying this assumption (`condemning this culprit'), which he called `Outcome Independence' (i.e. accepting its negation, Outcome Dependence) would secure a `peaceful coexistence' between quantum non-locality and relativity theory. 

My discussion will proceed in two main stages: first, Section \ref{preamble2}, and then Sections \ref{intro} to \ref{pi?}. The stages are linked by a common theme, viz. that Outcome Independence, and so also its negation, Outcome Dependence, use a merely {\em schematic} notion of `outcome'; (and similarly, their contrasted notion of `parameter' is schematic). This will mean that Outcome Dependence does not itself give the detailed physical account that peaceful coexistence needs (Section \ref{preamble2}). But this negative verdict prompts a positive project: to examine a detailed physical account---I take that of Adrian Kent---and assess whether Outcome Dependence holds in it. I do this in Sections \ref{intro} and \ref{od}. This project then prompts, in Section \ref{pi?}, a  final discussion of whether the obvious `alternative culprit'---the condition Shimony called `Parameter Independence'---holds in Kent's proposal. The main message of this final discussion will be to underline the importance of theorems by  Colbeck and Renner (made rigorous by Landsman), and by Leegwater.

The rest of this Introduction will spell out this plan in more detail.

Thus I begin in Section \ref{preamble2} by urging that our predicament is not as fortunate as Shimony hoped. Outcome Dependence does not---at least by itself---secure peaceful coexistence, for someone (such as Shimony and myself) seeking a `realist' and `one-world' interpretation of quantum theory. For the schematic notion of `outcome' in Outcome Dependence leads inevitably into the quantum measurement problem. So if one rejects solving this problem by adopting some kind of `instrumentalism', or by saving `realism' with some kind of `many worlds' solution, one needs a realist one-world---and relativistic---solution. Once given such a solution, one can then define `outcome' (unschematically!) and assess whether Outcome Independence fails.

So the gist of Section \ref{preamble2} is scepticism about Shimony's hope: forgive me, Abner! But I note that in his final years, he himself came to doubt the proposal (2009: 489; 2009a: Section 7, (1)). It is not just that he was a long-standing advocate of some process of dynamical reduction, i.e. of non-unitary evolution of isolated quantum systems.\footnote{\label{DRP}{For the dynamical reduction programme, cf. e.g. Bassi and Ghirardi (2003) and Pearle's essay (2009) in honour of Shimony. For Shimony's advocacy, cf. e.g. his (1990) and (2009a: Section 7, (2)). Excellent philosophical discussions include Myrvold (2002, 2003, 2017).}} He also cites Bell, who seems to have doubted the proposal. Although (so far as I know) Bell did not explicitly discuss Shimony's distinction between Outcome Independence and Parameter Independence (cf. Section \ref{preamble2} for definitions), he of course resisted making a fundamental interpretative distinction between outcomes (i.e. measurement results) and parameters (i.e. apparatus-settings). His viewpoint was that a pointer-reading (outcome) and a knob-setting (parameter) are both macrophysical facts, and so surely on equal terms, as regards whether a curious (i.e. unscreenable-off) correlation between examples of them at spacelike separation violates relativity theory. As he might put it: `surely Nature, in her causal structure, does not care whether a macrophysical fact is `controllable' by humans, in the way a knob-setting, but not a pointer-reading,  is---or at least seems to be?' This viewpoint is especially clear in his last essays, `Against measurement' (2004: p. 213-7, 227-30) and `La nouvelle cuisine' (2004: p.  237-8, 244-6): Shimony cites this last reference.  

From Section \ref{intro} onwards, I turn to the positive project. Section \ref{intro} introduces my chosen realist and relativistic one-world interpretation, namely Kent's (2014, 2015, 2016). Indeed, my aim is in part simply to {\em advertise} Kent's proposal. For I think philosophers' discussions of the interpretation of quantum theory, especially the measurement problem, focus too much on the `usual suspects', especially dynamical reduction, `many worlds' and the pilot-wave.\footnote{My (2015) is another such effort. But there are several other proposals deserving more attention from philosophers, such as those of Allahverdyan et al. (2013), and Landsman and co-authors (2013, 2013a, 2017 Chapters 10.1-3, 11.4).}    Then in Section \ref{od}, I use the fact that Kent's proposed beable---Bell's jargon for preferred quantity, whose extra values solve the measurement problem---makes precise, and unschematic, the idea of an outcome, to investigate whether his interpretation satisfies Outcome Dependence. 

The situation will be interestingly analogous to that for the pilot-wave theory. It is well-known to satisfy Outcome Independence at the `micro-level', i.e. the level of its hidden variables (particles' positions), while recovering Outcome Dependence at the observable level of experimental statistics by averaging over the hidden variables. I will urge that despite Kent's proposal being otherwise very different from the pilot-wave theory, the situation is analogous: Outcome Independence at the `micro-level', and Outcome Dependence at the observable level.	

This verdict prompts the question: what about the other much-discussed locality condition or `possible culprit'---the condition Shimony called `Parameter Independence'? Does it hold in Kent's proposal? This question is hard to answer, and must wait for another occasion. But in Section \ref{pi?}, I briefly address it. I will make two main points, both arising from some recent theorems. The first is about Kent, the second about peaceful coexistence. (I owe the first to discussion with Guido Bacciagaluppi and Gijs Leegwater: to whom my thanks.) \\
\indent \indent  \indent First: Theorems by Colbeck and Renner, and by Leegwater, give us powerful tools for addressing our question about Parameter Independence.  For they say (roughly speaking!) that, under some apparently natural assumptions: any theory that supplements orthodox quantum theory must  violate either a `no conspiracy' assumption, or Parameter Independence. It is clear, even allowing for rough speaking and for the unmentioned assumptions, that this result is important for understanding quantum theory, quite apart from evaluating Kent's proposal. After all: as is well-known, Bell was prompted to prove his first non-locality theorem by his awareness that the pilot-wave theory was non-local in the way we now call `Parameter Dependent', so that he naturally asked himself whether any supplementation of quantum theory had to be non-local.\footnote{As is also well-known, he asked himself this question in print, in the closing paragraph of his ground-breaking first paper on hidden variables: a paper which, due to an editorial oversight at {\em Reviews of Modern Physics}, was only published in 1966, i.e. after he had given a `Yes' answer to the question for deterministic hidden variables, in his 1964 proof of a Bell inequality.} Since then, we have learnt---again, following Bell's lead---to prove Bell inequalities for stochastic rather than deterministic hidden variables, and to distinguish subtly different `locality' assumptions: in particular, we have learnt, following Shimony, to distinguish between Outcome Dependence and Parameter Dependence. 

So the Colbeck, Renner and Leegwater theorems are remarkable---one might say, ironic---for leading us back, after all these years, to see Parameter Independence, which Bell long ago saw as violated by the pilot-wave theory, as {\em the} central locality notion---the notion that, on pain of some `conspiracy', any supplementation of quantum theory must deny. So here again, my aim will be in part  to {\em advertise} the theorems to philosophers---just as I wish to advertise Kent's proposal. 

So the obvious question arising here is whether Kent's proposal is again (as for Outcome Independence) analogous to pilot-wave theory: which obeys `no conspiracy' and of course violates Parameter Independence. At first sight, it seems that the answer is Yes, despite Kent's significant differences from pilot-wave theory. (These differences go well beyond his proposal being relativistic. For example, it does not use position as its preferred quantity (beable); and it invokes a final condition of its quantity, rather than, as in the pilot-wave theory, an initial condition.) But we will see that on reflection, the analogy falters. For Kent's proposal may violate `no conspiracy': though as I shall stress, `conspiracy'  is an unfair label, since there is nothing conspiratorial (or suspicious or `spooky') about the violation. So the upshot will be that, despite these theorems, Kent's proposal may yet obey Parameter Independence, even while supplementing orthodox quantum theory---and solving the measurement problem!

\indent \indent \indent My second, and final, point will turn on the fact that these theorems say that Parameter Independence and `no conspiracy' lead to `unsupplemented' quantum theory.  (Leegwater's theorem is especially impressive, in being free of assumptions beyond these two.) This means in effect, that they make some kind of peaceful coexistence mandatory for someone who, like Shimony, endorses Parameter Independence.  \\

\noindent To sum up this Introduction:--- The upshot of the paper will be twofold:\\
\indent (1): On the one hand, Shimony's hope for peaceful coexistence is alive and well. Indeed, recent theorems in a sense make it `mandatory'. \\
\indent (2): But on the other hand: peaceful coexistence needs more than a judicious or subtle choice of which assumption of a Bell theorem (which `locality condition') to deny. It will probably need no less than an agreed relativistic solution to the quantum measurement problem. And in seeking such a solution---in particular, in developing Kent's solution---we need to bear in mind whatever constraints on the solution are implied by general theorems like those of Colbeck, Renner and Leegwater.

\section{Does Outcome Dependence secure `Peaceful Coexistence'?}\label{preamble2}
In this Section, I will (i) introduce Shimony's proposal that Outcome Dependence promises peaceful coexistence (Section \ref{preamble2.1}); then (ii) report the details of the proposal (Section \ref{preamble2.2}), and (iii) give some reasons for scepticism about this promise (Section \ref{preamble2.3}). 

\subsection{Shimony's proposal}\label{preamble2.1}
Shimony proposed (1984, p. 131-136; 1986, 146-154; following Jarrett (1984)) that denying Outcome Independence led, or at least promised to lead, to some sort of `peaceful coexistence' between quantum nonlocality and relativity theory. This proposal was based on Jarrett's insight:  that the assumption that had hitherto been the main one in proofs of Bell inequalities (then usually called `factorizability' or `conditional stochastic independence') is a conjunction of two conditions---so that one could deny one, but not the other. (Details in Section \ref{preamble2.2}.) Shimony labelled them `Outcome Independence' (OI) and `Parameter Independence' (PI). 

Both assumptions concern the traditional two-wing Bell experiment (but can be generalized to set-ups with three or more wings/`parties'). Roughly speaking: Outcome Independence says that, conditional on sufficient information (including the choice of the two quantities to be measured), the outcome in one wing is stochastically independent of the outcome in the other wing; while Parameter Independence says that the outcome in one wing is (conditional on sufficient information: including the quantity chosen in that wing, and the outcome in the other wing) stochastically independent of the choice of quantity to be measured in the other wing. (So here, the quantity chosen is dubbed `parameter',  though `setting' would be clearer.) 

Thus Jarrett and Shimony proposed that in the light of Bell inequalities being violated, it is OI, not PI, that we should deny. It will be convenient to have labels for the negations of these assumptions: so we speak of `Parameter Dependence' (PD), and `Outcome Dependence' (OD). \\

\noindent Jarrett's and Shimony's reason for denying OI instead of PI, and saying that OD combined with PI led to, or at least promised, peaceful coexistence with relativity theory centred around the ideas that:\\
\indent \indent (i): since experimenters could choose parameters i.e. settings, PD would enable one experimenter to signal to the other her choice of parameter, which for  spacelike-related regions would amount to superluminal signalling; while on the other hand\\
\indent \indent (ii): experimenters could not choose (nor, apparently: influence) which outcome occurred, so that OD did not threaten superluminal signalling: it instead reflected, or at least suggested, a `holism' of the quantum correlations---which one might well hope relativity could  accommodate.\footnote{From a large subsequent literature, let me pick out just: an early collection, Cushing and McMullin (1989), Jarrett's essay in honour of Shimony (2009), Myrvold's recent (2016) statement of a position similar to Shimony's---and a more sceptical position of my own (2007: especially Sections 3.1 and 3.3, pp. 825-832, 846-851).}

This viewpoint was also supported by analysing which of the conditions, OI and PI, are obeyed by the two `obvious'	theories one might consider: on the one hand,  orthodox quantum theory; and on the other, its well-known rival---the pilot-wave theory. In the metaphor of the court-room: it was supported by these two theories' verdicts about these conditions.

The overall point here is that one can consider different theories' verdicts, because the conditions are  {\em schematic}. For they are equations about probabilities, with notations for outcomes, for apparatus-settings (i.e. quantity-choices, `parameters'), and a `hidden variable' or `complete state' of the pair: they say nothing physical about outcomes etc. So given a theory, it is a matter of judgment exactly which of its notions to take the schema's notations to refer to, so as to get a verdict on whether the condition is obeyed. But for these two theories, applied to the two-wing Bell experiment, there are very natural judgments about the interpretation of the schema's notations. And applying these, we find:\\
\indent \indent(i): orthodox quantum theory obeys PI, but not OI;\\ 
\indent \indent(ii): the pilot-wave theory obeys OI, but not PI.\\
Section \ref{preamble2.2} gives details. Then Section \ref{preamble2.3} urges scepticism.

\subsection{The conditions and their diverse verdicts}\label{preamble2.2}
\subsubsection{Parameter independence and outcome independence}\label{PIOI}
Consider stochastic models of the usual two-wing Bell experiment. We represent the two possible choices of measurement on the left (L) wing by $a_1, a_2$; and 
on the right (R) wing, by $b_1, b_2$. The idea is that a complete state (``hidden variable'') $\lambda \in \Lambda$  encodes all the factors that influence the measurement outcomes that are settled before the particles  enter 
the apparatuses, and that are therefore not causally or stochastically dependent on the measurement choices. So  $\lambda$ specifies probabilities for outcomes $\pm 1$  of the various single and joint  measurements:
\be
pr_{\lambda, a_i}(\pm 1) \; , \; pr_{\lambda,  b_j}(\pm 1) \; , \; {\rm{and}} \;\;\; 
pr_{\lambda, a_i, b_j}(\pm 1 \& \pm 1) \; ; i,j = 1,2.
\label{atlambda}
\ee
We also represent  outcomes by $A_i, B_i, i = 1,2$, where $A_i = \pm 1$ is the event that 
measuring $a_i$ yields $\pm 1$.  We  will also use $x$ as a variable over $a_1, a_2$;  $X$ as 
a variable over $A_1, A_2$ and their negations (i.e., outcomes $\mp 1$); and for the right wing, we 
similarly use $y$ and $Y$.

Observable probabilities are predicted by averaging over $\lambda$. For example, the 
observable  left wing single probability for $A_1 = + 1$ is:
\be\label{recoverbyaverage}
pr(A_1 = + 1)  := \int_{\Lambda} \; pr_{\lambda, a_1}(+ 1) \; d \rho \; .
\ee
Here, the use of the same measure $\rho$ irrespective of which quantity, e.g. $a_1$ vs. $a_2$, is chosen to be measured, encodes a locality assumption. Namely, that there is no correlation between (i) the causal factors influencing which quantity, e.g. $a_1$ vs. $a_2$, is measured, and (ii) the causal factors influencing which value of $\lambda$ is realized. 

This assumption seems very reasonable, especially if one takes the causal factors (i) to be localized in the wings of the experiment, and the causal factors (ii) to be localized in the central source of the emitted particle-pairs. One thinks: would it not be a {\em conspiracy} if there was a correlation between such surely disparate causal factors? So we shall grant this assumption without demur---for {\em most} of this paper. But at the end of Section \ref{crl}, we shall see that it can be denied for wholly unproblematic reasons, if $\lambda$ encodes facts about a final boundary condition, and thereby can encode information in that boundary condition about traces (records) of earlier choices of measurement. (So it would be misleading to think of such a $\lambda$ as localized in the central source.)  Besides: since the notion of a `hidden variable' $\lambda$ is schematic (as stressed at the start of Section \ref{preamble}), considering such a $\lambda$ is wholly legitimate.  On the other hand, a `reassurance':  invoking such a final boundary condition does not necessarily imply that any probability measure over the final boundary conditions must depend on which quantity has been previously measured.

In any case, the assumption of locality used in Bell's theorem, that is traditionally most focussed on, is: `factorizability' or `conditional stochastic independence'. This says: the joint probabilities prescribed by each value of $\lambda$ factorize into the corresponding single probabilities.\footnote{This condition is found in, for example: Bell's 1971 paper (2004: p. 36-38), Clauser and Horne (1974: eq. (2'), p. 528), and Shimony (2009a: Section 2 eq. (10), and Section 4 eq. (37)). The facts that this condition (i) occurs in both the Bell, and the Clauser and Horne, papers, and (ii) suffices, together with our earlier `no conspiracy' assumption (that $\rho$ in eq. \ref{recoverbyaverage} is independent of which quantity is measured), for a Bell inquality, were first clarified by Shimony, Horne, Clauser and Bell in a famous 1976 exchange. It is reprinted as Bell et al. (1985), and as Chapter 12 of Shimony (1993).} In our notation:
\be
\forall \lambda; \forall x,y; \forall X, Y = \pm 1: \;\;
pr_{\lambda, x,y}(X \& Y) =  pr_{\lambda, x}(X) \cdot pr_{\lambda,  y}(Y) \; .
\label{facby}
\ee

Jarrett's and Shimony's main formal point is  that eq. \ref{facby} is the conjunction of two apparently disparate independence conditions for a probability of a ``local'' i.e. ``this-wing'' result.\footnote{Jarrett (1984) is the most thorough early source for this `conjunction' point. But we should recall van Fraassen's brief but masterly exposition, which dubs Parameter Independence `Hidden Locality', and Outcome Independence `Causality' (1982: principles IV and III, respectively, p. 31). Incidentally, Shimony himself came to agree that `parameter' was too general a word for `distant setting'. His (2009a: equations 8 and 9) replaces the label `Parameter Independence' by `Remote Context Independence', and correspondingly `Outcome Independence' by `Remote Outcome Independence'. But I shall stick to using the established labels.} The first is, roughly speaking, independence from the measurement  choice in the other wing; called 
`Parameter Independence' (PI) (where `parameter' means `apparatus-setting'):
\be
\forall \lambda; x,y; X, Y = \pm 1: \;\;
pr_{\lambda, x}(X) = pr_{\lambda, x,y}(X) :=   pr_{\lambda, x,y}(X \& Y) + pr_{\lambda, 
x,y}(X \& \neg Y) \; ;
\label{pi}
\ee
and similarly for R-probabilities. The second condition is, roughly, independence from the outcome obtained in the other wing: `Outcome Independence' (OI):
\be
\forall \lambda, x,y; X, Y = \pm 1: \;\;
pr_{\lambda, x,y}(X \& Y) =  pr_{\lambda, x,y}(X) \cdot pr_{\lambda, x,y}(Y) \; .
\label{oi}
\ee

Then Bell's theorem states that any stochastic  model obeying eq. \ref{facby}, or equivalently, eq. \ref{pi} and \ref{oi} is committed to a Bell inequality governing certain combinations of probabilities: which is experimentally violated.  On the other hand, quantum theory is not  committed to such an inequality---and its predictions for these combinations of probabilities are confirmed.\footnote{Shimony (2009a: Sections 3 to 5) surveys experimental aspects. Fine recent discussions of the contents of the various versions of the theorem include Brown and Timpson (2016), Wiseman and Cavalcanti (2017).}

\subsubsection{The orthodox verdict: the `no signalling theorem'}\label{odox}
Given that Bell inequalities are violated by quantum theory and by experiment: the usual verdict is that OI is false. In particular, quantum theory obeys PI but not OI. More precisely: let us exploit the schematic nature of Section \ref{PIOI}'s conditions, and so put a quantum mechanical state for each $\lambda$, and take the probabilities at a given $\lambda$ to be given by the orthodox Born rule. Then we infer that:---\\
 
\indent \indent (a): PI holds. It is now a statement of the orthodox quantum no-signalling theorem (e.g. Ghirardi et al. (1980), Shimony (1984: 134-6),  Redhead (1987: Section 4.6, 113-116)). This theorem says that single-wing probabilities are not affected by any distant-wing {\em non-selective} measurement (i.e. measurement with no outcome selected). To prove it, the measurement process is often modelled as a projective or POVM measurement, i.e. with the ``projection postulate''. Thus in non-selective  projective measurement of a quantity $Q$ with pure discrete spectrum and spectral decomposition $Q = \Sigma \; q_n \Pi_n$, the initial density matrix $\rho$ is changed by measurement to $\Sigma \; \Pi_n \rho \Pi_n$. Then the theorem is immediately proven in the density matrix formalism, using the cyclicity of trace and the commutation of the L- and R-quantities (cf. Jordan 1983).\\
\indent \indent (b): OI fails, as a mathematical triviality except in the special case of the quantum state being a product state, which of course makes the relevant probabilities factorize. (In terms of density matrices: for any projectors $\Pi_L, \Pi_R$ representing outcomes on the left and right respectively, a product state $\rho_L \otimes \rho_R$ gives tr($\Pi_L \otimes \Pi_R \; . \; \rho_L \otimes \rho_R$) = tr($\Pi_L \rho_L$)tr($\Pi_R \rho_R$).) Of course, as Shimony and Jarrett admit, and much of the literature (cf. footnotes 5 to 8) stresses, this mathematical triviality should not blind us to the intuitive plausibility of factorizability, eq. \ref{facby}. One feels that, given the measurement choices (parameters): if $\lambda$ is the {\em complete} state of the pair, the probability conditional on it of a L-outcome must be unaffected by conditioning on further information about the R-outcome. It is from this intuition that the mysteriousness of---and historically, the surprise at---violations of Bell inequalities, arises. 

\subsubsection{The heterodox verdict: the  pilot-wave}\label{hdox}
 If we turn to the pilot-wave theory (e.g. Bohm and Hiley (1992), Bricmont (2016), Holland (1993)), then the orthodox quantum verdicts, (a) and (b), in Section \ref{odox} are {\em reversed}, once we identify the ``hidden variable'' $\lambda$ occurring in PI  and OI with what in the pilot-wave theory, one naturally considers the complete, or total, state. In the most-studied versions of pilot-wave theory, the preferred quantity (beable) whose extra values solve the measurement problem, is the position of point-particles.  So the natural notion of complete state is: the conjunction of (or: the ordered pair comprising) the quantum state (especially: the wave-function on configuration space) and the particles' possessed positions.
 
 This reversal of verdicts is normally stated for the non-relativistic pilot-wave theory, and its description of experiments that use Stern-Gerlach magnets to measure spin; and I will follow this. (But it also holds for relativistic versions, which retain an absolute simultaneity structure.). These descriptions take the wave-packet of one of the two particles, say the left particle, to be incident on a bifurcation plane across the magnet: the interaction with the magnetic field then splits the wave packet in two, the two halves being swept away from the plane---and the point-particle gets swept along within whichever `half-packet' it happens to be in. That is; it gets swept away from the plane without crossing it. So the pilot-wave theory makes precise the two possible outcomes---spin being `up' or `down' in the direction concerned---in as straightforward a way as you could wish for: in the point-particle being, indeed, `up' or `down' relative to the plane. (Cf. e.g. Dewdney et al. (1987: Sections 3-5, p. 4721-4730), Holland (1993, Sections 11.2-11.3, p. 465-476), Barrett (1999, 127-132), Bricmont (2016, Section 5.14, 141-150).)
 
Thus the situation is:---\\

\indent (a): PI fails. In each individual run of the experiment, there is action-at-a-distance; or, phrased more cautiously: instantaneous functional dependence of the value of a quantity `here', on the choice of a measurement setting at a distant location. The reason, in short, is that point-particles' possessed positions evolve according to the guidance equation---which for an entangled state of two particles, makes the velocity of each particle instantaneously dependent on the position of the other. In more detail: the momentum of each particle, $i = 1,2$ is given by ${\bf p}_i = \nabla_i S$, where $S \equiv S({\bf x}_1, {\bf x}_2)$ is the phase of the wave function in configuration space. This means that the possessed position of particle 1 (respectively, 2) contributes to {\em where} in configuration space the gradient is taken, for determining the momentum of particle 2 (respectively 1). So in a Bell experiment, the position of, for example, the R-particle, swept upwards from its bifurcation plane, contributes to determining the momentum, and so the later position, of the L-particle.  The orthodox quantum probabilities, and in particular the no-signalling theorem, are then recovered at the observable level, by averaging over the ``hidden variables'', i.e. the possessed positions, using the Born-rule distribution. (And more generally: the much-celebrated empirical equivalence with orthodox quantum theory is obtained by such averaging. But if another 
distribution is used, signalling {\em would} be possible. Besides, this point generalizes to other deterministic hidden variable theories that reproduce quantum theoretic statistics by a `quantum equilibrium' distribution of hidden variables;  (Valentini 2002, 2002a).)\\
\indent (b): OI holds. The reason, in short, is that Outcome Independence is trivial for a deterministic theory. Given the measurement choices (parameters), and a state rich enough to determine an L-outcome, conditioning on a distant R-outcome gives no further information. And of course, the pilot-wave theory, with $\lambda$ taken to contain not just the quantum state but also the possessed positions (or corresponding ``beables'' in other versions), {\em is} deterministic. (Agreed, there are subtleties about this last statement: (thanks to Bryan Roberts for stressing this). The existence and uniqueness of solutions for the combined Schr\"{o}dinger and guidance equations is guaranteed only under certain, albeit reasonable, conditions. But I shall set these subtleties aside: Berndl (1996) is a review, and details are in Berndl et al. (1995).)

\subsection{Too easy?}\label{preamble2.3}
Taken together, the verdicts in Sections \ref{odox} and \ref{hdox} seem to support the idea that OD, with PI, makes for peaceful coexistence between quantum theory and relativity. For orthodox quantum theory seems `at peace' with relativity, thanks to its no-signalling theorem; while the pilot-wave theory being `at war' is explicitly shown by its violating PI---a `state of war' that its obeying OI does not calm.\footnote{Similarly, some authors (Howard 1989, Teller 1989) suggested that the moral of OD was that quantum theory exhibited a species of holism or non-separability. Cf. Morganti (2009: Section 2) as an example of the ongoing philosophical discussion. But as the sequel indicates, I concur with Henson (2013: p. 1021-1028, Section 3.1-3.4) that this moral does not, as Henson puts it, `relieve the problem of Bell's theorem'.} 

But on examination, this support is questionable. It is not just that, as we have stressed, the conditions OD etc. are schematic. More specifically:\\
\indent \indent (1): Note that the no-signalling theorem uses only commutation of the two quantities that get measured. Nothing is assumed about the spacetime location of the measurements. Indeed, the theorem is often presented (e.g. Redhead 1987, p. 113-116; Bohm and Hiley 1992, p. 139-140) in a wholly non-relativistic quantum formalism that, ``notoriously'', allows superluminal propagation: in particular, wave-packets with an initially compact spatial support spreading instantaneously. So, as some authors (e.g. Muller 1999, p. 242) point out: for this formalism, the no-signalling theorem is really a `coincidence', since nothing in the conceptual framework of the formalism suggests a non-selective measurement must be forbidden from affecting the statistics of a distant measurement.\footnote{This point echoes Bell's viewpoint, mentioned in Section \ref{preamble}. But perhaps one should say, so as to reflect one's hope of reconciling the quantum with relativity: not `coincidence', but `manna from Heaven'.}\\
\indent \indent (2): Mentioning outcomes, and saying that they cannot be influenced (`controlled' as Shimony puts it) puts one face-to-face with the measurement problem: how does quantum theory represent a definite experimental outcome? Or perhaps better: how {\em can} it? Or: how {\em should} it? This problem was of course repeatedly pressed by Bell in his condemnation of orthodox quantum theory's `shifty split' (i.e. its vagueness about the quantum-classical transition), and its shifty replacement of the `and' of superposition by the `or' of an ignorance-interpretable mixture (2004: p. 93-4, 117-8, 155-6, 213-7, 245-6). Even setting aside all issues about  quantum nonlocality, this problem still has no agreed answer---and the best of authorities continue to press the problem (e.g. Isham (1995, Sections 8.5, 9.4), Leggett (2002), Ghirardi (2009), Landsman (2017, Chapter 11)). Besides: considering  quantum nonlocality  only  aggravates the problem. For we have no agreed relativistic description of quantum measurement  processes: in particular, no agreed relativistic formulation of the `collapse of the wave-packet' (whether the collapse is treated as a fundamental physical process, or as in some way effective or even subjective). And we have no consensus about how to obtain one. Again, various authors have pressed the problem.  Among discussions of how OD in particular bears on these issues, earlier work includes Butterfield (1992: Section 7, p. 72f.), Clifton and Jones (1993), Muller (1999); and recent work includes Norsen (2009, 2011) and Henson (2013). All these authors are sceptical that OD secures `peaceful coexistence'.

Agreed, I have stated the measurement problem---as do Bell and the other authors cited---in terms that set aside solutions that are `instrumentalist' rather than `realist', or that  save `realism' with some kind of `many worlds' solution.\footnote{Fine recent versions of `instrumentalism' include Friederich (2015, Chapters 7-10) and Healey (2017: Chapter 4, especially Section 4.6 pp. 72-74, and Chapter 10, especially Section 10.6, pp. 179-183: building on his previous (2012, 2013, 2014)). For `many worlds', Wallace (2012) is nowadays the {\em locus classicus}.} And I frankly admit to hoping, as Shimony did, for a realist one-world---and relativistic---solution. Hence my positive project, from Section \ref{intro} onwards, to focus on one such proposal. \\

\noindent To sum up this Section: we cannot expect peaceful coexistence to be readily established: in particular, not just by appealing to OD. The word `outcome'  leads to the measurement problem; and correlatively, so does the contrast made by OI and PI between outcomes and parameters. The apparent neatness of the contrast in the equations of OI and PI belies  controversial issues about what the quantum state represents, and how measurement processes unfold---especially in a relativistic spacetime.

\section{Kent's proposal for a realist one-world Lorentz-invariant interpretation}\label{intro}
This Section presents Kent's proposal (2014, 2015, 2016): first its strategy (Section \ref{strategy}), then its details (Section \ref{details}). The details involve three stages, of which the third is the main one (Section \ref{3stages}). Then we can see how Kent recovers both the empirical success of orthodox quantum theory, and  a single actual quasiclassical history (Sections \ref{recover} and \ref{2constraints}).

\subsection{The strategy}\label{strategy}
We imagine, to begin with, that we are given a Lorentz-invariant quantum theory defined on Minkowski spacetime, which is able to rigorously describe interactions, in particular measurements, and which describes the total system as evolving unitarily. 

Agreed:  we in fact---notoriously---do not have a rigorous formulation of an empirically adequate Lorentz-invariant quantum theory describing interactions. (Not even such a theory of the basic interactions between microphysical entities like electrons and photons, let alone measurement interactions.) But Kent's proposal for how we should augment such a theory can be understood, and  assessed, without knowing all the details of such a theory. For although Kent's proposal must refer to interactions, he endorses (and indeed uses) the conventional though non-rigorous  Lorentz-invariant physics of interactions. And there is every reason to think that what he postulates additionally, about the values of a preferred physical quantity (traditionally `observable'; but better: `beable') and their probabilities, will {\em not} conflict with this conventional physics: nor with a rigorous formulation of it, were we ever to have such.\footnote{\label{asymp}{As we will see, the spacetime need not be Minkowski: any appropriately  causally well-behaved spacetime will do. There is a more substantial constraint, arising from his appeal to a final boundary condition: viz. that there should be a well-defined limit to a sequence of probability distributions,  associated with a sequence of successively later and later spacelike hypersurfaces. But as Kent discusses, there is good reason to think this will be satisfied in favoured cosmological models.}}

The aim is to augment this theory in a precise way so as to give a realist one-world Lorentz-invariant interpretation of quantum theory (modulo the issues just mentioned about describing interactions both rigorously and empirically adequately).\footnote{Remark: Kent's focus is on the measurement problem, which he prefers to call {\em the reality problem}, `since few physicists now believe that the fundamental laws of nature involve measuring devices per se or that progress can be made by analysing them' (2014, 012107-1). Hence the title of his paper. I agree with his preference for `reality problem' over `measurement problem': but I will keep to the traditional term, reflecting his wider concern by talking about seeking an `interpretation of quantum theory'.}

Kent augments this theory by specifying---all in a suitably Lorentz-invariant way:\\
\indent \indent (i) an appropriate physical quantity (traditionally `observable'; but better: `beable');  \\
\indent  \indent (ii) possible temporal sequences of values for it; and \\
\indent \indent (iii) probabilities for those sequences:\\
such that the one real world corresponds to one such sequence (taken together, of course, with the total history of the orthodox unitary evolution of the quantum state, according to the given theory).

In other words:  Kent specifies within the given theory a beable, and thereby a sample space of its possible values, and histories of values; on which he then defines a probability measure (in terms of a sequence of final conditions, each given a conventional Born-rule probability), so as to recover both:\\
  \indent   \indent   (a) the successful standard quantum description of microphysics (in particular: the principle of superposition and Lorentz-invariance), and \\
  \indent   \indent   (b) the successful standard classical  description of macrophysics, i.e. the emergence of a quasiclassical history. \\
Thus each total history of the universe is given fundamentally by the conjunction of \\
  \indent   \indent (a') the history throughout time of the quantum state, which evolves unitarily and Lorentz-invariantly (so (a') might be called the `orthodox part of the history'---which Everettians claim is the whole history); and \\
  \indent   \indent (b') the history throughout time of the beable's actual possessed values: i.e.  one specific trajectory through the sample space.\\

{\em Remark}:  So far, the broad `natural philosophy' of the proposal seems to be like the pilot-wave theory: a unitary quantum evolution is conjoined with the history of the actual possessed values of the beable. Besides, as in the pilot-wave theory: there is no dynamical back-reaction from the actual possessed value to the quantum state and-or its evolution. But as we will see in Section \ref{details}, there are substantial differences. Overall, the traditional language of `hidden variables' sits ill with Kent's ideas (hence his use of `beable'), and it will be clearest to forget it until we return to Outcome Dependence in Section \ref{ODPIInToy}. More specifically, the principal differences from pilot-wave theory are: \\
  \indent   \indent  (i) fundamental Lorentz-invariance (of course); \\
  \indent   \indent  (ii) a different choice of the beable (not: position), and a different  prescription for how its actual possessed value evolves (not: a deterministic `guidance equation'); 	\\
  \indent   \indent  (iii) a different prescription for the probabilities of the various possible possessed values (not: the Born-rule applied to the initial values of the beable, and proven equivariant for the unitary evolution).

\subsection{The details}\label{details}
The details of Kent's proposals vary between his three papers. In particular, the first includes (in its Section II) an extended presentation of an analogous proposal for a non-relativistic spacetime; and its relativistic proposal differs significantly from the second and third papers. The second paper gives toy models of photons scattering off a massive quantum system whose initial wave-function is an archetypal `two hump' superposition (in one spatial dimension) of `being on the left' and `being on the right'. In these models, the photons are treated as point-like objects propagating along light-like curves, and interacting with the massive quantum system by reflection. The third paper's models are more realistic: Kent treats the photons as well as the massive systems quantum mechanically, using the formalism of photon wave mechanics. But in all three papers, his proposal secures his desired result: that there is, under appropriate circumstances, an `effective collapse' onto one or other location---cf. (b) and (b') at the end of Section \ref{strategy}. 

I will concentrate on the {\em second} paper: discussing first, the detailed proposal in three stages (Section \ref{3stages}), and then the recovery of a quasiclassical history (Sections \ref{recover} and \ref{2constraints}).  

\subsubsection{Three stages}\label{3stages}
The main idea of the choice of beable is, in a word, that it should be mass-energy. And the main idea about the probabilities of its various values unfolds in three stages:\\
\indent (i) to consider the orthodox Born-rule probabilities, prescribed by the quantum state, for mass-energy's possible values at a suitable late time (i.e. at all points on a suitable late spacelike hypersurface), though there is no assumption that an actual physical measurement is made anywhere on this hypersurface; and then \\
\indent (ii) to take one possible distribution of mass-energy on this hypersurface as a final boundary condition; and \\
\indent (iii) to evaluate probabilities for values of mass-energy at an earlier spacetime point by (a) using the quantum state (for the appropriate time), {\em but also} (b) conditioning on the final boundary condition given by (ii).\\

More precisely:--- Kent first recalls that the Tomonaga-Schwinger formalism enables us, given a quantum state $| \psi_0 \rangle$ prescribed on some initial spacelike hypersurface $S_0$, to define formally the evolved state $| \psi_S \rangle$ on any hypersurface $S$ in the future of $S_0$ via a unitary operator $U_{S_0S}$. This formalism enables him to fulfil stage (i) above. Thus we are to envisage a world in which physics plays out between two hypersurfaces $S_0$ and $S$, and a quantum state is given on $S_0$. We consider the local mass-energy density operators $T_S(x)$ for $x \in S$. (So as usual, $T_S(x) =  T_{\mu \nu}(x)\eta^{\mu}(x)\eta^{\nu}(x)$  where $T_{\mu \nu}(x)$ is the stress-energy tensor at $x$, and $\eta$ is the future-directed unit 4-vector orthogonal to $S$ at $x$.) Then the quantum state $| \psi_S \rangle \equiv U_{S_0S} | \psi_0 \rangle$ prescribes orthodox Born-rule probabilities for the various possible distributions of values of all these operators. But of course, there is no need to suppose that a joint measurement of these operators in fact occurs on $S$.\footnote{After expounding stages (i)-(iii), Kent discusses taking $S$ ever later in spacetime, i.e. letting $S$ go to future infinity; and therefore his proposal needs the probability distributions associated to ever later $S$ to have an appropriate limit: cf. footnote \ref{asymp}. But I shall not discuss this aspect.}

Kent then proposes that one possible mass-energy distribution $t_S(x)$ on $S$ is randomly selected, using the Born-rule probability distribution prescribed by $| \psi_S \rangle$. That is: physical reality includes one such distribution. This is stage (ii). 

As for stage (iii), i.e. proposing probabilities for the beable mass-energy at a spacetime point $y$ between $S_0$ and $S$, Kent proposes that these should be calculated conditionally on---not the whole final boundary condition---but only on that part of it that lies outside the future light-cone of the spacetime point $y$. The effect of this, as we will see in Section \ref{recover}, is that there can be an `effective collapse' of appropriate superpositions of values of mass-energy at intermediate points $y$ (thus securing the desired definiteness of macroscopic quantities), thanks to photons scattering differently off the components of the superposition and then later registering differently on  part of the surface $S$, and so contributing to the final boundary condition. Note that this collapse is by construction Lorentz-invariant: roughly speaking, a `collapse along the light-cone'. 

But before discussing that, here is a summary of stages (i) to (iii), in Kent's own words (2015, Section 2 (a)):
 
\begin{quote}
We wish to define a generalized expectation value for the stress-energy tensor at a point $y$ between $S_0$ and $S$,
using post-selected final data $t_S(x)$ on $S$. More precisely ... we will use the post-selected data $t_S(x)$ for all points $x \in S$ outside the future light cone of $y$, {\em and only for those points.} [Kent labels the set of all these points $S^1(y)$.] 

In words, our recipe is to take the expectation value for $T_{\mu \nu}(y)$ given that the initial state was $| \psi_0 \rangle$ on $S_0$, conditioned on the measurement outcomes for $T_S(x)$ being $t_S(x)$ for $x$ outside the future light cone of $y$. So, for any given point $y$, our calculation ignores the outcomes $t_S(x)$ for $x$ inside the future light cone of $y$.

[Kent then defines a sequence $S_i(y)$ of spacelike hypersurfaces that: (i) include  almost all of $S^1(y)$, i.e. almost all of the part of $S$ outside the future light cone of $y$; and (ii) include $y$; and (iii) as $i \rightarrow \infty$, get ever closer to that part of the future light cone of $y$ that lies to the past of $S$.] 

Now for any of the $S_i(y)$, we can consider the Born rule probability distribution of outcomes of joint measurements
of $T_S(x)$ (for all $x \in S \cap S_i(y)$) and of $T_{\mu \nu}(y)$. These are calculated in the standard way, taking the initial state $| \psi_0 \rangle$ on $S_0$, unitarily evolving to $S_i(y)$, and applying the measurement postulate there. . . . By taking the limit as $i \rightarrow \infty$ we obtain a joint probability density function $P(t_S(x),t_{\mu \nu}(y))$ [for $x \in S^1(y)$]. From this, we can calculate conditional probabilities and conditional expectations for $t_{\mu \nu}(y)$, conditioned on any set of outcomes for $t_S(x)$ (for $x \in S^1(y)$),  in the standard way.

Our mathematical description of reality, in a hypothetical world in which physics takes place only between $S_0$
and $S$ and in which the outcomes $t_S(x)$ were randomly selected, is then given by the set of conditional expectations $\langle t_{\mu \nu}(y) \rangle$ for each $y$ between $S_0$
and $S$, calculated as above. We stress that the calculations for the beables $\langle t_{\mu \nu}(y) \rangle$
at each point $y$ all use the same final outcome data $t_S(x)$. However, different subsets of these data are used in these calculations: for each $y$, the relevant subset is $\{ t_S(x) : x \in S^1(y) \}$.
\end{quote}

\subsubsection{Recovering quantum theory's empirical success, and a quasiclassical history}\label{recover}
Kent now proceeds to recover both:\\
\indent \indent (a) the empirical success of standard quantum theory in microphysics; and\\
\indent \indent (b) a single quasiclassical history in macrophysics: (cf. (a) and (b) at the end of Section \ref{strategy}). 

To do so, he relies
on the existence of ``environmental'' `particles, or wave packets, or field perturbations, that travel at light speed' (2015, Sec 1). But as we shall see, Kent's appeal to the ``environment'' is judiciously different from a {\em mistaken} (though all too common!) appeal to decoherence: in short, he does not make the error of thinking that an improper mixture is ignorance-interpretable.

Kent's main idea here can be usefully divided into two parts. The first part can be briefly stated; the second will occupy the rest of this Subsection. It will require discussion of decoherence; and this discussion will lead to Kent's needing two constraints to hold good---discussed in Section \ref{2constraints}. (Kent argues, by considering toy models (2015, Section 3; 2016, Examples 1-5), that there is good reason to think the constraints do hold good.)\\

{\em The first part}:--- The first part just assumes our universe has such lightlike propagations (I will say `photons', for short) and then applies Section \ref{3stages} to them. So the first part says that these photons, propagating from some point $y$ in the spacetime, arrive on the later spacelike hypersurface $S$, and:\\
\indent (i) by registering there, these photons contribute to the actual values (outcomes of Kent's notional measurements) $t_S(x)$ for $x \in S^1(y)$, i.e. to the actual mass-energy density distribution on the hypersurface $S$; \\
\indent (ii) by being correlated, according to the unitarily evolving quantum state, with other degrees of freedom at $y$, these photons function as {\em records} of those degrees of freedom; in particular, the relevant subset  $\{ t_S(x) : x \in S^1(y) \}$ of the actual values determines the beable $\langle t_{\mu \nu}(y) \rangle$ at the point $y$, i.e. the expectation of $T_{\mu \nu}(y)$ conditional on $\{ t_S(x) : x \in S^1(y) \}$---cf. the end of Section \ref{3stages}.\\

{\em The second part}:--- The second part is less general, and  explicitly directed at recovering (a) and (b) above. So it is, inevitably, less systematic than the first part: Kent supports it with some analyses of toy models (2015, Section 3; 2016, Examples 1-5). To understand this second part, I propose that we think of it as a reconstrual, within Kent's framework, of the  insight (nowadays universally accepted) of decoherence theory, that:\\
\indent (i): when a quantum system is {\em very} well isolated, so that the very fast, efficient and ubiquitous process of decoherence, arising when a quantum system interacts with its environment (e.g. photons, or air molecules), can be avoided or at least postponed: the interference terms, that are characteristic of the system being in a superposition rather than a mixture, will persist and characteristic quantum phenomena (like the iconic interference patterns in the two-slit experiment) will occur; whereas\\
\indent (ii): when the quantum system is {\em not} well isolated from its environment, decoherence will rapidly  ``diffuse'' the interference terms out into the environment: so that the system's reduced state is a mixture---and accordingly, one is tempted to say that a component of the mixture represents a quasiclassical history (more precisely: an instantaneous slice, or member, of such a history). 
 
 But note: Kent's postulate of a specific beable and its actual values $t_S(x)$, and so of actual values $\langle t_{\mu \nu}(y) \rangle$ make this insight, about (i) vs. (ii), play out differently, as regards conceptual aspects (though not  numerical aspects), from the way it usually plays out when  decoherence is  invoked in discussions of the measurement problem. To better understand Kent's proposal, it will be helpful to spell out these contrasts: (and as presaged at the end of Section \ref{strategy}, Kent's proposal will be similar in some respects to the pilot-wave theory). Spelling out these contrasts will also prepare us for Section \ref{od}'s discussion of Outcome Dependence. \\

{\em Contrasts with `decoherence as usual'}:--- So recall the idea of decoherence: plausible Hamiltonians for the interaction between a quantum system that is comparatively massive---paradigmatically, the pointer of an apparatus---and another system or systems that is/are comparatively light---paradigmatically, the air molecules or photons scattering off the pointer---imply that after the interaction, the reduced state (i.e. density matrix) of the pointer is nearly diagonal in a variable that is a collective variable encoding information about mass and position. Thus in some models, it is nearly diagonal in the centre of mass of the pointer. (The reason for the implication is, broadly speaking, that interaction Hamiltonians are local in position.)  This result prompts two striking and much-discussed suggestions in relation to the measurement problem. Both will give a contrast with Kent's proposal (and with the pilot-wave theory).

First: Notice that these models give a dynamical explanation of the salience, or `selection', of a quantity such as the centre of mass of the pointer. The quantity is not given a special role---e.g., that of always having a value---by some general {\em ab initio} postulate: it is just made salient, or selected, by the nature of the interaction in question. This marks a contrast with Kent, who postulates a special role for mass-energy density and other components of stress-energy. (And as mentioned at the end of Section \ref{strategy}: it marks a contrast with the pilot-wave theory which postulates (in its usual version) a special role for the position of point-particles.)   

Second: The pointer being in a mixture for a quantity such as the centre of mass suggests the measurement problem is solved! It is tempting to exclaim: surely, it represents the desired mixture, as against superposition, of macroscopically distinguishable configurations! As is nowadays well known, this is a chimera. In d'Espagnat's (1976, Chapter 6.2)  hallowed terminology: the mixture is `improper' i.e. it is {\em} not ignorance-interpretable---as it would have to be, in order to solve the measurement problem at one fell swoop, along the lines envisaged.\footnote{Though much could be said about this topic, this is not the place: except for two short points, the first historical and the second conceptual. (1): Though I  duly cited d'Espagnat's clear and influential presentation of the point, and nowadays the decoherence literature often says it clearly (e.g. Zeh, Joos et al. (2003, p. 36, 43); Janssen (2008, Sections 1.2.2, 3.3.2)), it is humbling to recall that Schr\"{o}dinger already was clear on it in his amazing 1935 papers: cf. especially the analogy with a school examination (1935, Section 13, p. 335 f.). (2): Although I thus condemn this `one fell swoop' solution (joining, of course, much wiser authorities including Bell): I of course agree that the result here---obtaining an improper mixture nearly diagonal in a quantity you `want' to have definite values---can form an important part of a principled, and clear-headed, solution to the measurement problem. The obvious example is the modal interpretation, in its early versions from the mid-1990s (cf. Dieks and Vermaas 1998): which, roughly speaking, proposed as a postulate, going beyond quantum theory, that the eigenprojections of any system's density matrix have definite values.} 

But Kent makes no such error. The postulate of a randomly selected final condition, i.e. the postulated fact of one specific mass-energy distribution $t_S(x)$ on the later spacelike hypersurface $S$, gives a single, definite value to the conditional expectation $\langle t_{\mu \nu}(y) \rangle$. The point here is again similar to the situation for the pilot-wave theory. It also makes no such error. For according to it (in the usual version), just one component of the density matrix, in the fundamental position representation, has the relevant point-particles in its support: thus giving a single, definite position. (And again, for both Kent and the pilot-wave theory: there is no back-reaction from the actual, randomly selected, value of the beable to the evolving universal quantum state.)\\ 

To sum up this Subsection: Kent proposes we can achieve the two goals (a) and (b)---we can recover both (a) the empirical success of standard quantum theory in microphysics, and (b) a single quasiclassical history in macrophysics---and we can do so in a Lorentz-invariant way. To do so, we invoke---not: position as a beable---but mass-energy as a beable at a late time (i.e. spacelike hypersurface $S$) together with its orthodox Born-rule correlations to the expectation values of components of stress-energy at earlier spacetime points. Lorentz-invariance is respected by  appropriately restricting  which part of the entire distribution $t_S(x)$ on $S$ is conditioned on, for determining the expectation value at an earlier spacetime point $y$.

\subsubsection{The recovery needs two constraints to be satisfied}\label{2constraints}
Combining Section \ref{3stages}'s discussion of Kent's three stages, and Section \ref{recover}'s discussion,  we can now see that Kent's proposal  depends on two constraints holding good. I will call them, ($\alpha$) and ($\beta$). They correspond, respectively, to the goals (a), `accurate microphysics', and (b), `a single macrohistory',  listed at the start of Section \ref{recover} (and repeated in its final paragraph). But I should of course note that those goals (and so also the constraints below) are connected, e.g. because we use the statistics of macrophysical pointers, i.e. facts about the quasiclassical world, to confirm quantum theory's description of microphysics.\footnote{For clarity and simplicity, I will again suppress the need to let the late spacelike hypersurface $S$ go to future infinity. Cf. footnotes 11 and 13.}\\ 

\indent ($\alpha$): Quantum theory's empirically successful descriptions of microphysical phenomena, e.g. interference patterns in the two-slit experiment: \\
\indent \indent  \indent  (i) are recorded in (the expectation values of) mass-energy and other components of stress-energy at appropriate points $y$ in spacetime, e.g. in the positions of massive pointers in front of a calibrated dial; and\\
\indent \indent  \indent (ii) are thus recorded with statistics that are close to the orthodox Born-rule probabilities prescribed by the quantum textbook,  i.e. by the quantum state ascribed in the standard manner to the measured system (so that indeed, orthodox applications of quantum theory are vindicated, by Kent's lights); and \\
\indent \indent  \indent (iii) these textbook probabilities are equal, or close enough, to the probabilities prescribed in Kent's stage (iii) of Section \ref{3stages}. Namely: equal or close to a probability derived by combining:\\
\indent \indent  \indent \indent \indent  [i] the  correlation (in simple cases: strict or near-strict correlation) of the relevant component(s) of stress-energy at $y$ with appropriate features of the final condition $t_S(x)$ on $S$; with \\
\indent \indent  \indent \indent \indent [ii] the orthodox Born-rule probability of those appropriate features of  $t_S(x)$: i.e. the probability  prescribed by  $| \psi_S \rangle$,  the unitary time-evolute on $S$ of the universe's initial state $| \psi_0 \rangle$ on $S_0$. \\

Three comments on this constraint, ($\alpha$), before I turn to the second constraint.

First: Note that all three of ($\alpha$)'s clauses (i)-(iii) are needed in order to link Kent's proposed beables, and his proposed probabilities of their values, with quantum theory's empirical success, and with how we know it (i.e. how we confirm quantum theory by collecting experimental statistics). \\

Second: In ($\alpha$)'s clause (iii), we are to consider some single actual value (final condition: outcome of a notional measurement) $t_S(x)$: conditioning on which defines the beable $\langle t_{\mu \nu}(y) \rangle$ at the earlier spacetime point $y$; (cf. Section \ref{3stages}'s exposition of Kent's stage (iii)). The idea is: the state of the environmental particles (for short: photons) encodes a value of the beable at $y$, and later on this gives a contribution to (i.e. a non-zero component of) the quantum state $| \psi_S \rangle$; and, we can suppose, this contribution survives in the randomly selected actual final condition $t_S(x)$ (i.e. $t_S(x)$ actually includes this contribution). Thus the random selection on $S$ actually including this contribution, combined with this contribution's strict correlation to the value of the beable at the earlier point $y$, {\em makes it true} that the beable has that value at $y$. Cf. also the discussion of constraint ($\beta$) below.\footnote{\label{formallike}{So in terms of the formalism: Kent's idea is {\em like} an appeal to one `branch' of the state of an environmental particle, i.e. one component of its improper mixture, to pick out as factual the corresponding (strictly correlated) component of the improper mixture of the system of interest. But only `like', not `identical with'!  As I stressed  in Section \ref{recover}'s {\em Contrasts with `decoherence as usual'}: the difference is that Kent {\em avoids} the error of assuming an improper mixture is ignorance-interpretable. He knowingly postulates the beables, and their probabilities, that secure an actual quasiclassical history, while avoiding this error.}} \\ 

Third: Let us ask: How plausible is it that ($\alpha$) holds good? Of course, a conclusive assessment is not possible, in the present state of knowledge. After all, recall from the beginning of Section \ref{strategy} that, quite apart from Kent's proposals, we lack a rigorous Lorentz-invariant quantum field theoretic account of interactions. So we have no rigorous relativistic theory  of quantum measurement, and we cannot now rigorously test clauses (i) and (ii), even though they concern only Born-rule probabilities prescribed by the quantum textbook, i.e. by the standard quantum state of the measured system. A rigorous test of clause (iii) is even more difficult:  it corresponds to Kent's stage (iii)  of Section \ref{3stages}, i.e. the calculational algorithm that needs appropriate correlations between beables to be encoded  in the universal state $| \psi_S \rangle$. So testing ($\alpha$)'s clause (iii) amounts to combining, in a realistic, relativistic setting:\\
\indent \indent [i] the sorts of ideas and formalism used in the physics of measurement-processes and decoherence (and we must again note sadly that hitherto, this physics is almost exclusively studied in a non-relativistic setting ...); with \\
\indent \indent  [ii] the sorts of ideas and formalism used in pilot-wave theory's discussions of effective quantum states (and conditional wave-functions) whereby one justifies attributing a quantum state (i.e. a wave-function, not merely an improper mixture) to a subsystem of the universe, in terms of both the universal state and the actual values of the subsystem's beables: ideas and formalism that would then be adapted to Kent's postulated beables.\\
Clearly, in the present state of knowledge, the best we can do by way of testing ($\alpha$) is to look at various toy models of measurement: for example, with photons scattering off some massive object (thought of as a pointer) and later registering on a hypersurface. This, Kent proceeds to do (2015, Section 3; 2016, Examples 1-5). These models give detail to the ideas  I have sketched here, and in my second comment above. Cf. also the second constraint that Kent needs, ($\beta$)---to which I now turn.\\

\indent ($\beta$):  The actual single quasiclassical history in macrophysical terms---the sequence through time of the values of countless macrophysical quantities (such as whether the centre of mass of Erwin's cat's tail is above the floor and moving (`alive'!) or on the floor and still (`dead')), with the sequence obeying approximately classical laws of motion, and so attesting to the emergent validity of classical  physics---is {\em picked out}  by the actual mass-energy distribution $t_S(x)$ on the late hypersurface $S$. Here, `picked out' means: the expectation values of components of stress-energy at the various points $y$ throughout spacetime, that encode the actual single quasiclassical history, have strict, or nearly strict, Born-rule correlations, according to the calculational algorithm in Kent's stage (iii), with the actual mass-energy distribution $t_S(x)$ on the late hypersurface $S$. More precisely, so as to respect the light cone structure: `picked out' means that each  expectation value at a point $y$ has such a correlation with the actual mass-energy distribution $t_S(x)$ in that part of $S$ that is outside $y$'s future light cone. \\

Finally, we need to address a question about this constraint, ($\beta$): a question which relates back to my second and third comments on constraint ($\alpha$). One is bound to ask:\\
\indent \indent What, if anything, does Kent's proposal need to say about   the fact that  we cannot  now know the  actual mass-energy distribution $t_S(x)$ on the hypersurface $S$; and relatedly, what does it need to say about the fact that we cannot now (or ever) know more than a tiny fraction of the actual single quasiclassical history? \\
This question---these two pools of ignorance---prompts two remarks: the first conceptual, the second empirical.

\indent (1): {\em How to represent a definite perception?}:--- Agreed, these pools of ignorance do not cause any immediate problem for Kent. Nothing in the exposition above (or in Kent's papers) requires us (or anyone) to know the  actual mass-energy distribution $t_S(x)$, or even anything substantive about it. Nor, correspondingly, are we or anyone required to know facts about the actual single quasiclassical history. But . . . there is an issue here, about how we should conceive the representation {\em in that quasiclassical history} of our knowledge of it---partial, indeed tiny, though that knowledge no doubt is. Thus suppose that in the actual history, the centre of mass of Erwin's cat's tail is above the floor and moving (i.e. the infernal device did not kill the cat---the actual world is better than it might have been ...), and Erwin knows this, since he sees the tail vertical and moving. Then this definite perception, and the knowledge it engenders, are also part of the actual definite quasiclassical history---and so presumably, Kent proposes that they are represented by appropriate values of appropriate beables. I do not  wish here to foist on Kent an account of the relation between mental and physical states, or even require that he should have some such  account. But clearly, there is an issue to consider. Namely: do definite perceptions (and their consequent states of belief and knowledge) correspond to values of components of stress-energy, i.e. of the {\em same} beable that Kent has already proposed as sufficient to secure a definite {\em inanimate} macroscopic realm? Or are the subtleties of the mental/physical nexus such that  they correspond to values of some other beable?\footnote{Again, there is an interesting comparison with the pilot-wave theory. For my question to Kent is the analogue of a question sometimes raised about the pilot-wave theory. Namely: does the `psycho-physical parallelism'  (von Neumann 1932, Chapter VI.1 p. 418f.) between some mental states, such as states of perceptual knowledge, and some physical states of our sensory organs (e.g. depression of a touch receptor in my fingertip, or photons impinging on my retina) mean that a perception being definite---one way rather than another---involves a point-particle being in one wave-packet  rather than another?  (For example, cf. Brown and Wallace (2005: Section 7, p. 533-537).) But I think that on this topic, Kent's situation is {\em much}	 more comfortable than the pilot-wave theorist's. For according to modern psychophysics, it is much more plausible that mental states being one way rather than another correspond to (i) values of components of stress-energy at locations in the brain being one way rather another, than to (ii) point-particles being in one location  rather than another.}  

(2):   {\em How to assess the constraint?}:--- But these pools of ignorance {\em do} cause a problem for efforts to assess whether this constraint ($\beta$) for `recovering a quasiclassical history' in fact holds good or fails. If we are to assess ($\beta$), despite our ignorance, we will have to somehow simplify and-or idealize,  as regards both (1) the final  actual mass-energy distribution and (2) the details of the actual single quasiclassical history. For both (1) and (2), the obvious strategy is to look at a toy model, including very simple idealizing assumptions about both the environmental particles and the quasiclassical history. This is of course what Kent does (2015, Section 3; 2016, Examples 1-5). For example, his simpler models assume that `photons' scatter off macroscopic systems without any recoil by the latter; and that the quasiclassical history consists just of the locations of one or more macroscopic massive quantum systems that have zero self-Hamiltonian. 

Thus his simplest example, which uses just one spatial dimension, goes roughly as follows. I will develop this example in Section \ref{KentToy}, so as to assess Outcome Dependence.\\
\indent \indent  (i) A  macroscopic massive quantum system with zero self-Hamiltonian is initially stationary in a two-peak superposition of two locations, call them `left' and `right'; \\
\indent \indent (ii) it then scatters a photon off one peak---{\em and} the other, i.e. the photon's quantum state after the interaction is itself a two-peak wave-packet (more precisely: an improper mixture with two equi-weighted components) thus encoding both possible locations of reflection;\\
\indent \indent (iii)  later on, the photon registers on the  hypersurface $S$ where---we can suppose---the randomly selected actual final condition encodes that the photon's location is such as to record that it had earlier scattered off the massive system's {\em left} peak, not its right one---so that\\
\indent \indent (iv) applying Kent's calculational algorithm (especially stage (iii) of Section \ref{3stages}), the macroscopic massive quantum system is localized on the left: that is, the beable $\langle t_{\mu \nu}(y) \rangle$ is substantially non-zero for $y =$ `left', and is zero, or close to zero, for  $y =$ `right'.\\

So much by way of briefly expounding Kent's proposal. Obviously, there is a great deal one could  explore here: for example, the detail of Kent's toy models, and his suggestions for developing them, or for varying the postulates so as to get `cousin' theories that could differ empirically from quantum theory.  But I will now confine myself to assessing the fate, in Kent's proposal, of the conditions, Outcome Independence and Parameter Independence. As announced in Section \ref{preamble}, there will be an interesting analogy with the pilot-wave theory: similar verdicts on Outcome Independence, and---I believe!---on Parameter Independence. And we will see a connection with some recent no-go, i.e. `no hidden-variable' theorems.
 
\section{Outcome Independence?}\label{od}
I now investigate whether Kent's proposal satisfies Outcome Dependence. I will argue that the situation is analogous to that for the pilot-wave theory. As we saw in Section \ref{hdox}, the pilot-wave theory satisfies Outcome Independence. That is: it is satisfied at the `micro-level', i.e. the level of hidden variables (particles' positions). But by averaging over the hidden variables using, indeed, the orthodox Born-rule probability distribution, the theory recovers orthodox quantum theory's Outcome Dependence  at the observable level of experimental statistics (cf. Section \ref{odox}, (b)). 

Thus I will argue that despite Kent's proposal being otherwise very different from the pilot-wave theory, the situation is analogous: \\
\indent \indent (i) Outcome Independence at the `micro-level', i.e. using probabilities conditioned on a specific value of Kent's beable, i.e. specific values of the mass-energy distribution on (appropriate parts of) the late hypersurface $S$, $\{ t_S(x): x \in S \}$; while on the other hand, there is:\\ 
\indent \indent (ii) Outcome Dependence at the observable level, after averaging over these values, using the orthodox Born-rule probabilities prescribed by the quantum state $| \psi_S \rangle \equiv U_{S_0S} | \psi_0 \rangle$ on $S$.	

To make this argument, I need to adapt the ideas of Kent's toy models, as summarized at the end of Section \ref{2constraints}, to a Bell  experiment.   I will first quote Kent's own presentation of his first toy model (Section \ref{KentToy}). Here I will emphasize that Kent's invocation of a final boundary condition is conceptually unproblematic.  Then I adapt the ideas to (a very simple model of) a Bell experiment (Section \ref{adapt}). Then in the last Subsection (Section \ref{ODPIInToy}), I conclude, as announced in (i) and (ii) above, that Kent and the pilot-wave theory give similar verdicts about whether Outcome Independence is obeyed: Yes at the micro-level, No at the observable level. So this last Subsection will return us to the language of `hidden variables', which we set aside in order to expound Kent's ideas; (cf. the {\em Remark} at the end of Section \ref{strategy}).

\subsection{Kent's first toy model}\label{KentToy}
Kent presents his first toy model as follows (2015, Section 3). The algebra in this quotation---in particular, the arguments in the $\delta$-functions describing the positions of the point-like photons---can be checked by looking at the spacetime diagram (p. 27: thanks to Bryan Roberts): where $X$ is the position coordinate of the photon.
\begin{quote}
We consider a toy
version of ``semi-relativistic" quantum theory, in which a
non-relativistic system interacts with a small number of
``photons".  We treat the photons as following light-like path
segments.  We model their interactions with the system as bounces,
which alter the trajectory of the photon.  For simplicity, we neglect
the effect of these interactions on the non-relativistic system, and
also neglect its wave function spread and self-interaction, so that in
isolation its Hamiltonian $H_{\rm sys} = 0$.  We simplify further by
working in one spatial dimension, and we take $c=1$.

We suppose that the initial state of the system is a superposition of two separate localized states, $\psi^{\rm sys}_0 =   a \psi^{\rm sys}_1 + b \psi^{\rm sys}_2$.   Here $ |a|^2 + |b|^2 = 1$ and the $\psi^{\rm sys}_i$ are states localized around the points $x = x_i$, with $x_2 > x_1$. 
For example, the $\psi^{\rm sys}_i$ could be taken to be Gaussians (but recall that we are neglecting changes in their width over time).   
We take $ | x_1 - x_2 | $ to be large compared to the regions over
which the wave functions are non-negligible.   We thus have a crude
model of a superposition of two well separated beams, or of a macroscopic
object in a superposition of two macroscopically separated states.  

We suppose that the environment consists of a single photon, initially
unentangled with the system.    It is initially propagating rightwards from the direction $x = - \infty$, so that in the absence of any interaction it would reach $x= x_1 $ 
at $t=t_1$ and $x= x_2$ at $t= t_2 = t_1 + (x_2 - x_1 )$.

We take the photon-system interaction to have the effect of 
instantaneously reversing the photon's direction of travel, while
leaving the system unaffected.   (As noted above, we neglect the
effect on the system: this violates conservation of momentum but 
simplifies the overall picture.)    

Thus, for $t<t_1$, the state of the photon-system combination in our
model is 
$$
\delta (X - x_1 - t + t_1) ( a \psi^{\rm sys}_1 ( Y ) + b \psi^{\rm sys}_2 (Y) ) \, , 
$$
where $X,Y$ are the position coordinates for the photon and system respectively.  
	
For $t_1 < t < t_2$, the state is 
$$
\delta (X - x_1 + t - t_1 ) ( a \psi^{\rm sys}_1 (Y) ) + 
\delta (X - x_1 - t + t_1 ) (  b \psi^{\rm sys}_2 (Y) ) \, . 
$$

For $ t > t_2 $, the state is 
$$\delta (X - x_1 + t - t_1 ) ( a \psi^{\rm sys}_1 (Y) ) + 
\delta (X - x_2 + t - t_2 ) (  b \psi^{\rm sys}_2 (Y) ) \, . 
$$

The possible outcomes of a (fictitious) stress-energy measurement at a
late time $t = T \gg t_2$ are thus either finding the photon heading
along the first ray $ X = x_1 + t_1 - t$ and the system localized in
the support of $\psi^{\rm sys}_1$, or finding the photon heading along
the second ray $X= x_2 + t_2 - t$ and the system localized in the
support of $\psi^{\rm sys}_2$.

Suppose, for example, we consider a real world defined by the first
outcome.  Our rules for constructing the system's beables imply that,
for $t< 2 t_1 - t_2 $, and for $x = x_1$ or $x_2$, we condition on
none of these outcomes, since all of them correspond to observations
within the future light cone.  Up to this time, then, the mass density
beables for the system are distributed according to $ | \psi^{\rm sys} (Y)
|^2$, with a proportion $ | a |^2 $ localized around $Y= x_1$ and a
proportion $ | b |^2 $ localized around $Y= x_2$.

For $ 2 t_1 - t_2 < t < t_1 $, the observation of the photon on the first ray is outside the future light cone of the component of the system localized at $x_2$, but not outside the future light cone of the component localized at $x_1$.    This gives us mass density beables distributing a proportion $ | a |^2 $ of the total system mass around $x_1$, but zero mass density beables around $x_2$. 

For $ t > t_1 $, the observation is outside the future light cone of both localized components of the system.    This gives us mass density beables distributing the full system mass around $x_1$, and zero around $x_2$. 

In other words, in the picture given by the beables, the system is a
combination of two mass clouds with appropriate Born rule
weights initially, and ``collapses" to a single cloud containing the full mass after $ t > t_1$. 

\end{quote}

\noindent This quotation illustrates Kent's proposal well. To sum up: We suppose the  real world is defined by the first outcome. That is: the photon that entered from the left (flying rightwards) reflects from $x_1$ at $t_1$, and registers on the given  time slice $t = T$ (which is much later: $T \gg t_2 > t_1$). Given this outcome, there is zero-mass around $x_2$, in our frame, at all the times $t$ for which the  photon  hits (while flying leftwards) the  time slice $t = T$ outside the future light cone of $(x_2, t)$. That is: at all times $t$ later than $2t_1 - t_2$: a semi-infinite period. During the first part of this period---to be precise: for $t_1 > t > 2t_1 - t_2$--- the  photon  hitting the   $t = T$ time slice is still {\em inside}  the future light cone of $(x_1, t)$. It is only after $t_1$, i.e. when  $t > t_1$, that hitting the much later slice is outside the  future light cone of $(x_1, t)$. Thus the `collapse' to zero-mass around $x_2$ happens, in our frame, before the  `collapse' to full-mass around $x_1$ happens.

\begin{center}
\includegraphics[width=0.75\textwidth]{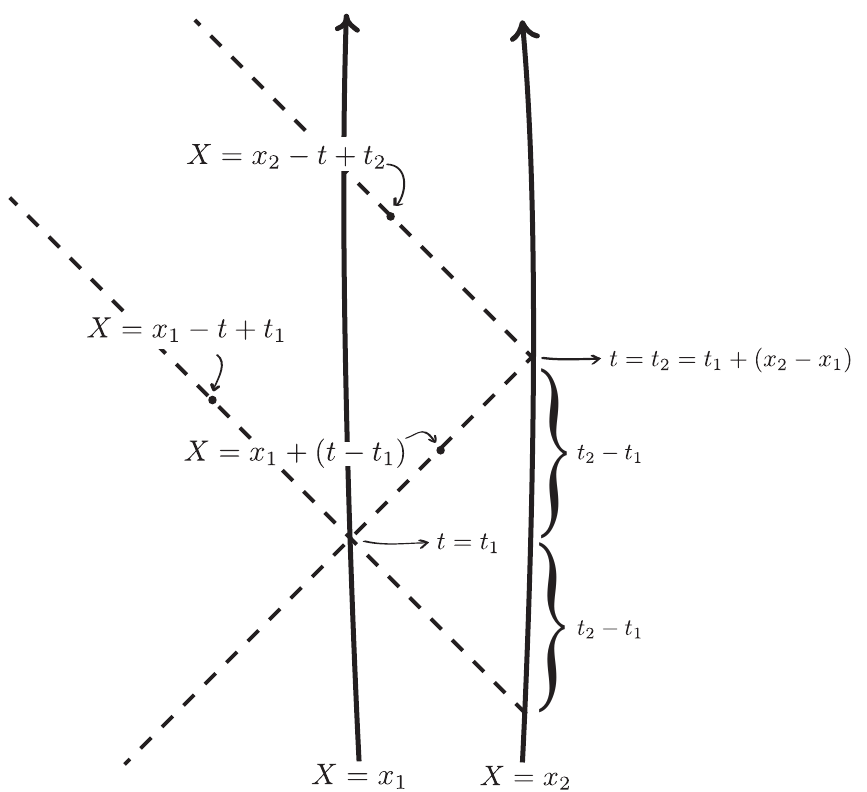}
\end{center}

Finally, we must {\em beware} of beguiling words! Thus it is tempting to say things like: the photon registering on the late spacelike hypersurface records that it reflected from one peak of the superposition, rather than the other. If such statements are read {\em without} their usual temporal connotations, they are indeed innocent: they suggest that there is (in a timeless or `block-universe' sense of `is')  an actual fact as to which reflection happens---and agreed, on Kent's proposal, there {\em is} such a fact. It is made true by the actual final condition: a final condition which is a matter of happenstance, of random selection (though not of an actual measurement). Thus in the preceding paragraph, my verbs like `registers', `hitting' and `happens',  are all to be read {\em without} temporal connotations: as what philosophers call `tenseless' verbs, despite their syntactic form being the same as present-tensed verbs.\footnote{There is nothing suspect about such tenseless verbs. They are not a philosophers' fiction or contrivance: the verbs in proverbs, e.g. `a stich in time saves nine',  and in pure mathematics, e.g. `1 + 1 = 2', are tenseless.} 

But beware: such statements (especially words like `registering', and `records that it reflected') normally {\em do} carry temporal, indeed causal, connotations. So they suggest---not just that there is an actual fact as to which reflection happens (where `happens' is tenseless!): which is true on Kent's proposal---but that:\\
\indent \indent (i) this actual fact is independent of the selection of the actual final condition; and even that \\
\indent \indent (ii) it {\em was} `made true', or `settled', before the time of the final condition.\\
And (i) and (ii) are, according to Kent, false. As I said  in Section \ref{2constraints}'s second comment on the constraint ($\alpha$): the randomly selected actual final condition {\em makes it true} that the beable has its value (the value it in fact has) at the earlier time. And in this last claim, the verbs are tenseless!

With this warning in mind, I submit that Kent's invoking a final boundary condition is conceptually unproblematic.


\subsection{A Kentian toy model of a Bell experiment }\label{adapt}

Let us now adapt the ideas of Section \ref{KentToy} to give a toy model of a Bell experiment. I will use the obvious analogy with pilot-wave descriptions of such experiments that use Stern-Gerlach magnets to measure spin. Recall from the summary in Section \ref{hdox} that the pilot-wave theory makes precise the two possible outcomes---spin being `up' or `down' in the direction concerned---in a straightforward way: by the point-particle being `up' or `down' relative to the bifurcation plane. 

Kent's proposal makes outcomes precise in a similar way. Agreed: his proposed beable is---not an always-existing (and continuously and deterministically evolving) point-particle position, but---roughly speaking, localized mass (or mass-energy) density. But this means that he can represent the two `latent' outcomes of a spin measurement by a superposition of two separate localized states, $\psi^{\rm sys}_0 =   a \psi^{\rm sys}_1 + b \psi^{\rm sys}_2$---just like the initial state of the massive system in Section \ref{KentToy}. And the occurrence of a single definite outcome, in the one actual world, is to be represented by a photon registering on the late spacelike hypersurface (i.e. by the random selection of the actual final condition) and so recording that it reflected from one peak rather than the other.\footnote{\label{renewwarn}{Here again, the words `registering', and `recording that it reflected', are to be understood tenselessly, in line with the warning at the end of Section \ref{KentToy}. That is: there is indeed an actual fact as to which reflection happens (tenselessly!). But this fact is {\em not} independent of, nor is it `made true' or `settled' before the time of, the actual final condition.}}   

To spell this toy model out a little, I assume that we again idealize by using just one spatial dimension. So let the locations $x_1$ and $x_2$,  as in Kent's model (so with $x_2 > x_1$), now be  in the left-wing of the experiment. The representations of the two possible outcomes of a  spin (or polarization) measurement on the massive quantum system entering the left wing (`the L-system') are the localization of mass density around $x_1$ and $x_2$ respectively. Again, the analogy is with how a Stern-Gerlach magnet makes the point-particle's position represent its spin. But I idealize by having just one spatial dimension, so that there is no bifurcation plane. And I will not try to represent alternative settings (parameters, components of spin) for the spin measurement: that will only become a topic in Section \ref{pi?}'s discussion of Parameter Independence.  

Of these two possible outcomes, one rather than the other occurs, in accordance with the actual final condition including a photon registering at a place on the late hypersurface $t=T$ that records a previous reflection off one peak rather than the other.
    
So much by way of describing the left-wing of the experiment. To describe the other wing, we assume there is also another massive system, the R-system, located far along the $x$-axis, and entangled with the the L-system. So: considering only spatial degrees of freedom (suppressing spin degrees of freedom), and ignoring how to treat the various possible settings (parameters) of measurement apparatuses, the initial joint quantum state can be written as (with $Y_L, Y_R$ for the spatial coordinates of the left- and right-systems, respectively):
\be 
a \psi_1(Y_L)\psi_4(Y_R) + b \psi_2(Y_L)\psi_3(Y_R)
\label{jointtoy}
\ee
where $x_3$ and $x_4$ are located far along the $x$-axis, i.e. $x_3 >> x_1, x_2$ and $x_4  >> x_1, x_2$, and with $x_3 < x_4$ ; and where each factor wave-function $\psi_i$ has support in a small neighbourhood of $x_i$. This state correlates  the two systems' positions: there is Born-rule probability $|a|^2$ of their being in their respective outer positions (i.e. $x_1$ and $x_4$) and Born-rule probability $|b|^2$ of their being in their respective inner positions (i.e. $x_2$ and $x_3$). (This anti-correlation---i.e. the fact that one particle gets localized at its left alternative iff the other gets localized at its right alternative---is of course a simple analogue of the anti-correlation of spin results for parallel settings, on the singlet state of two spin-half particles.)     

Now recall that Kent's first toy model, reviewed in Section \ref{KentToy}, had one photon that (i) initially propagated rightwards  from $x = - \infty$, but (ii) was supposed, by way of an example,  to register `left enough, soon enough' in the actual final condition so as to imply an earlier reflection at $x_1$ rather than $x_2$. 

So in the Bell experiment, we can imagine, in a similar way, two photons:\\
\indent (a) one initially propagating rightwards from $x = - \infty$, and as in Section \ref{KentToy}, later registering `left enough, soon enough' in the actual final condition so as to imply an earlier reflection at $x_1$ rather than $x_2$.\\
\indent (b) one initially propagating leftwards from $x = + \infty$, and later registering `right enough, soon enough' in the actual final condition so as to imply an earlier reflection at $x_4$ rather than $x_3$. \\
So (a) and (b) specify the joint quantum system's outcome as: each component system is localized in its outer position, i.e. the L-system gets localized at $x_1$ (its left alternative) and the R-system gets localized at $x_4$ (its right alternative).

So much by way of adapting the ideas of Kent's first toy model to give a (very!) toy model of a Bell experiment. Of course, other more complicated, less idealized models, are possible. But I have said enough to yield a verdict about Outcome Independence.

\subsection{Outcome Independence at the `micro-level', but not at the observable level}\label{ODPIInToy}
All the pieces are now in place. We only need to combine:\\
\indent \indent  (a): Section \ref{2constraints}'s constraint ($\alpha$), especially clause (iii), and constraint ($\beta$); with\\
\indent \indent (b): Kent's postulated beables making precise, viz. as localizations of mass-energy, the outcomes of measurements in Section \ref{adapt}'s toy model of a Bell experiment.\\
Combining (a) and (b), the analogy with the verdicts given by pilot-wave theory (Section \ref{hdox}) will be obvious.\\
 
 \subsubsection{The micro-level}\label{431micro}
Let us begin at the `micro-level'. That is: we fix attention on the actual (but of course unknown) final condition $t_S(x)$ on the late hypersurface $S$ in a universe where a Bell experiment is performed at a much earlier time, but after an initial hypersurface $S_0$.  As stressed in Section \ref{3stages}: there is no claim that any measurement is made anywhere on the hypersurface $S$. The idea is rather that the actual final condition is part of the one real world: a part that by its orthodox quantum  correlations (both strict and not strict) with earlier events, contributes to specifying the world---in particular, it specifies a quasiclassical history. Besides, it does so in a Lorentz-invariant way, thanks to Kent's calculational algorithm respecting the light-cone structure. 

More precisely, in light of footnotes 12 and 14:  There is not a single selected late hypersurface $S$; but rather, Kent postulates that:\\
\indent \indent (i) there is a well-defined limit to the distributions over all possible final conditions associated with an appropriate sequence of successively later hypersurfaces; and that \\
\indent \indent (ii) one element of the sample space on which the limiting distribution is defined---one `outcome' in the jargon of probability theory (not our jargon!)---is actual; and this `outcome', by its orthodox quantum correlations with various events throughout spacetime, specifies a quasiclassical history---including outcomes in {\em our} sense of macroscopic experimental results, such as pointer-readings.\\
But as in Section \ref{intro}, I will set aside this subtlety, and talk of an actual final condition on the hypersurface $S$, not the less intuitive `outcome' (ii) in a vast sample space. 

So we wish to consider probabilities about  events pertaining to the experiment, that are prescribed by the universal quantum state, but conditional on this actual final condition. In this endeavour, we will be guided by the evident analogy with the pilot-wave theory. Namely: Kent's actual final condition is like the pilot-wave theorist's actual possessed positions of all the point-particles; and we wish to calculate probabilities from, i.e. conditional on, both the orthodox quantum state and this extra `micro-level' information. Besides, the future-{\em and-past} determinism of unitary dynamics means that, although the pilot-wave theory usually bases its description of time-evolution on an initial quantum state and initial particle positions, it could instead use  the final quantum state and final particle positions---a `temporal direction' of description like Kent's. 

Thus, we recall the second paragraph of Section \ref{3stages}'s quotation from Kent: `our recipe is to take the expectation value for $T_{\mu \nu}(y)$ given that the initial state was $| \psi_0 \rangle$ on $S_0$, conditioned on the [notional] measurement outcomes for $T_S(x)$ being [the actual] $t_S(x)$ for $x$ outside the future light cone of $y$'. But we now need to amend the discussion so as to consider events,  not at one spacetime point intermediate between $S_0$ and $S$ (labelled $y$ in Section \ref{3stages}), but at four. 

So we replace the spacetime point labelled $y$, and its neighbourhood, by  four appropriately chosen spacetime points and the regions around each of them. Their spatial coordinates will be given in Section \ref{adapt}'s notation as  $Y_L= x_1, Y_L= x_2, Y_R= x_3, Y_R= x_4$ , where $Y_L (Y_R)$ is the position coordinate of the Bell experiment's L-system (R-system) respectively (cf. eq. \ref{jointtoy}). And their temporal coordinates are chosen to match when (one run of!) the experiment is in fact performed. (So one expects the two left, respectively two right, spacetime points to be nearly simultaneous; and to arrange the pair of left points to be spacelike separated from the pair of right points.) 

Let us suppose that in this run of the experiment the `outer', $x_1$ and $x_4$, outcomes in fact occur in the quasiclassical history specified by the actual  final condition $t_S(x)$. That is: $t_S(x)$ on $S$ encodes that  just after the run is completed, the L-system is  localized (tenseless!) around $Y_L= x_1$, and the R-system is  localized (again, tenseless) around $Y_R= x_4$; where this encoding is done by $t_S(x)$ describing photons registering in appropriate locations of $S$. To return to our example with one spatial dimension, described at the end of Section \ref{adapt}: the idea is that:\\
\indent \indent (a) a photon registers on $S$ so far to the left as to imply that it earlier reflected at $x_1$ rather than at $x_2$, while \\
\indent \indent (b) another photon registers on $S$ so far to the right as to imply that it earlier reflected at $x_4$ rather than at $x_3$.\\
So the idea is that if we condition orthodox quantum probabilities for experimental outcomes on all this (very rich!) information  in $t_S(x)$, the probabilities will become trivial, i.e. 0 or 1. And so they will factorize. On our supposition, we shall obtain probability 1 for each of the two `outer', $x_1$ and $x_4$, outcomes. 

To use the language of hidden variable theories: {\em the theory is deterministic}. That is: Kent's proposal, as spelt out in this toy model---with its strict correlations between where on $S$ a photon registers, and where the massive system has its mass-energy localized---is {\em past}-deterministic in the sense that in an individual run of the experiment, the later facts about photon registration, taken together with facts about the 	quantum state, imply with certainty  what the earlier outcomes were. So {\em Outcome Independence is satisfied.} Again, the situation is analogous to pilot-wave theory, with its familiar, present-to-future, kind of determinism: in that theory, the earlier facts about particle positions, taken together with facts about the quantum state, imply with certainty  what the later outcomes will be. Besides, as I noted five paragraphs above: the analogy can be strengthened. For the pilot-wave theory can instead use the final condition (of particle positions and quantum state), not the initial condition. \\

 \subsubsection{The observable level}\label{432oble}
I turn to the observable, i.e. experimental, level. You might say that Kent's proposal obeying Outcome Dependence (i.e. violation of eq. \ref{oi} of Section \ref{PIOI}) at this level is simply to be expected since, as we have seen  (especially in Section \ref{recover} and \ref{2constraints}), Kent's proposal is designed to match the  experimental probabilities of quantum theory---which obey Outcome Dependence. But this dismissal would be too quick, for two reasons. (Both of them are about getting a better understanding of Kent's proposal.) First: Kent gives a precise meaning to `outcome' in terms of localizations of mass-energy, and a calculational algorithm for probabilities of outcomes; so it is an important consistency check on his ideas to see that these lead to Outcome Dependence. Second: conducting this check provides an interesting comparison with the pilot-wave theory. 

So I return to the earlier discussion (at the end of Section \ref{2constraints}) about the fact that we of course do not know the final condition, and never will know it: and indeed, we will never know more than a tiny fraction of  the actual single quasiclassical history.  

I think Kent's proposal about this is natural; (compare clause (iii) of the first constraint, ($\alpha$), in Section \ref{2constraints}). Namely, he proposes that:\\
\indent \indent (1) one should average over the possible final mass-energy distributions $t_S(x)$ on  $S$---and do this averaging  with their Born-rule probabilities as prescribed by the final quantum state $| \psi_S \rangle$; and \\
\indent \indent (2) it is these averaged probabilities that are equal, or close enough, to the probabilities given by what I called the quantum textbook.\\
In other words:  clause (iii) of constraint $\alpha$ is interpreted as: these textbook probabilities are (nearly) equal to a mixture of probabilities  with (a) each component of the mixture conditioned on one of the various possible $t_S(x)$ (on the part of $S$ outside the relevant point's future light cone); and with (b) each component weighted with the orthodox  Born-rule probability, prescribed by $| \psi_S \rangle$, of its $t_S(x)$. To sum up: Kent proposes averaging over the possible final mass-energy distributions with their Born-rule probabilities, so as to recover the experimentally confirmed orthodox quantum probabilities. 

Again, there is an evident analogy with the pilot-wave theory. It also recovers these probabilities by averaging over the values of its hidden variable, and by doing this averaging with the values' Born-rule probabilities. But agreed, there is also a difference: we have a much more detailed understanding of how the  pilot-wave theory recovers orthodox quantum probabilities---indeed, so successfully that it is usually considered empirically equivalent to orthodoxy, at least for non-relativistic physics---than we do of how Kent's proposal does so.\footnote{This difference implies no criticism of Kent: the pilot-wave theory is long-established, and most expositions of its recovery of orthodoxy do not involve any cosmological, and so inevitably speculative, considerations of the kind Kent's proposal must tangle with.} Hence our answer to question (2) at the end of  Section \ref{2constraints}: that one must, as Kent does, investigate toy models. 

The upshot as regards Outcome Independence (eq. \ref{oi} of Section \ref{PIOI}) is clear. In so far as we are confident that Kent's proposal recovers orthodox quantum probabilities at the observable level, it of course satisfies Outcome Dependence at that level. Besides, since  Outcome Dependence is an inequation, not an equation:  even if the recovery is not exact, i.e. there are systematic differences, so that Kent's proposal is an empirical rival to orthodox quantum theory, not an interpretative addition to it: nevertheless, Kent's proposal probably satisfies Outcome Dependence at the observable level.

\section{Parameter Independence?}\label{pi?}
So much for Outcome Dependence. What about Parameter Independence: does it hold in Kent's proposal? 

The first point to make is that as we saw in Section \ref{ODPIInToy}, we should distinguish (as the pilot-wave theory does) between:\\
\indent \indent (i) the `micro-level', i.e. probabilities conditioned on specific final boundary conditions of Kent's kind; and \\
\indent \indent (ii) an observable level, obtained by averaging over these boundary conditions: which is expected to match, or  nearly match, orthodox quantum probabilities---though of course, this match should not just be assumed, but should be checked.

Assessing Parameter Independence, at either of these levels, is a substantive task: and not just because proving a universally quantified equation is harder than proving an existentially quantified inequation. Again, it is a matter of seeing how Kent's proposals for beables and their probabilities mesh with the formalism of standard quantum theory. Thus one needs  to explicitly model in a Kentian fashion both:\\
\indent \indent (a) different choices of apparatus-setting (parameter) on one wing (the right wing in eq. \ref{pi} of Section \ref{PIOI}), and \\
\indent \indent  (b)  non-selective measurement, (i.e. defining the left-wing marginal probability by summing probabilities of the various possible right-wing outcomes). \\
A Kentian model of (a) will involve photons scattering off the apparatus' knob that sets which quantity gets measured, and registering a long time later on the late hypersurface; so that the actual, randomly selected, final condition encodes the setting. Then in order to model (b), the probabilities dependent on reflection from ``which knob-setting'' will need to be combined with Section \ref{ODPIInToy}'s probabilities  dependent on reflection from ``which outcome, i.e. mass-energy lump''.

As stressed in Section \ref{PIOI} after eq. \ref{recoverbyaverage}, and at the end of Section \ref{KentToy}, there is nothing problematic or suspicious about a final condition encoding a knob-setting; nor about probabilities depending on such a setting. But writing down a model incorporating (a) and (b) {\em is} a substantive task. It is not obvious how such an explicit model would relate to standard quantum theory's no-signalling theorem, and more generally to the commutation of quantities. So I leave all this for another occasion!
 
But as announced in Section \ref{preamble}: recent theorems by Colbeck and Renner (made rigorous by Landsman), and by Leegwater, say (roughly speaking!) that, under some apparently natural extra assumptions: any theory that supplements orthodox quantum theory, in the sense of recovering orthodox Born-rule probabilities by averaging over other probabilities,  must  violate either a so-called `no-conspiracy' assumption, or Parameter Independence. (Again: `no-conspiracy' is an unfair label, since there need be nothing conspiratorial or problematic about a violation.) So Kent's proposal needs to be compared with these theorems, with a view to assessing which of their assumptions it satisfies, and which it violates (and maybe we will need to add: under which disambiguations). Besides,  as Section \ref{ODPIInToy}'s discussion brought out: Kent's proposal's probabilities at  the micro-level, i.e. probabilities conditioned on a specific final boundary condition, are {\em not} equal to orthodox Born-rule probabilities. So indeed, his proposal supplements quantum theory in the relevant sense; and so the proposal must violate one or more of the assumptions of these theorems. And since Leegwater's theorem seems to dispense with the need for Colbeck and Renner's extra assumptions, we infer that Kent's proposal must  violate either the no-conspiracy assumption, or Parameter Independence (at the micro-level), or both.  

Properly working out which it violates must wait for another occasion. I will just, in Section \ref{crl}, discuss the suggestion that it obeys the no-conspiracy assumption, and so {\em must} violate Parameter Independence. Then  in Section \ref{compel}, I make a final point, again based on these theorems: in effect, a peace-pipe to smoke with my dedicatee, Abner Shimony. Namely: these theorems, especially Leegwater's theorem, promote peaceful coexistence of the kind Shimony sought.       
 
\subsection{The theorems of Colbeck and Renner, and of Leegwater: Parameter Dependence at the micro-level?}\label{crl}
Colbeck and Renner claim to show that, under certain assumptions, `no extension of quantum theory can have improved predictive power' (2011, 2012). The idea here of `no improved predictive power' means, in the usual language of `hidden variables', that the hidden variables are otiose, or trivial. A bit more precisely: if the hidden variable `underpinning' of quantum theory is to recover orthodox Born-rule probabilities by probabilistic averaging over hidden variables using a measure $\mu_{\psi}$ dependent on the quantum state $\psi$, then (under certain assumptions) the hidden variables are trivial in the following sense. The probability prescribed by any hidden variable $\lambda$ for any outcome for a measurement of any quantity (and in any context of simultaneously measuring some other quantities) is the same as the orthodox Born-rule probability---except perhaps for a set of hidden variables $\lambda$ of zero $\mu_{\psi}$-measure. 

This is a stunning no-go result against a hidden variable underpinning of quantum probabilities, especially since the needed assumptions look natural enough; and indeed, Colbeck and Renner's proof-method is dazzlingly inventive. But their derivations are heuristic, rather than rigorous; and unsurprisingly, they prompted scrutiny, and even criticism. One concern was that their `freedom of choice' assumption (about experimenters being free to choose what to measure) in fact encoded Parameter Independence. Indeed, anyone familiar with the pilot-wave theory will suspect that Parameter Independence is `in there somewhere among'  Colbeck and Renner's assumptions. For in the pilot-wave theory, the hidden variables, i.e. the point-particles' positions, are certainly {\em not} otiose in the above sense, and they violate Parameter Independence (cf. Section \ref{hdox}). Agreed, this concern might be met by simply admitting Parameter Independence as an explicit assumption. 

But there were also other concerns about Colbeck and Renner's derivations: \\
\indent (a) other assumptions, and the proof itself, needed to be formulated more rigorously;\\
\indent (b) although these other assumptions (i) look natural enough, even when formulated rigorously, and (ii) are obeyed by quantum theory (in the sense of substituting the quantum state $\psi$ for the schematic hidden variable $\lambda$, just as we did in Section \ref{odox}), nevertheless: a hidden variable advocate might well doubt them---just as there is a noble precedent, viz. the pilot-wave theory, for denying Parameter Independence.

Of course, I cannot pursue all the topics that arise from the scrutiny of Colbeck and Renner's result. I will confine myself to reporting two main points: combining them will then lead directly to my conclusion about Kent.\\
\indent  First: Landsman has fully addressed concern (a) (while also emphasizing (b)). That is: he has given a mathematically rigorous statement of the assumptions, and a correspondingly rigorous proof of the theorem (2015; 2017, Chapter 6.6). \\
\indent Second: Leegwater (2016) has a significantly different but also rigorous theorem that is a `cousin' of the Colbeck and Renner theorem, as made rigorous by Landsman. Indeed, Leegwater manages to dispense with all the extra assumptions in (b) above, as clarified by Landsman, except for one: viz. the `no-conspiracy' assumption, that the measure used to average over the hidden variables so as to recover Born-rule probabilities---while of course it can depend on $\psi$, as above---must be independent of which quantity, or quantities, are (chosen to be) measured. (We already saw this assumption way back in Section \ref{PIOI}, in the discussion after eq. \ref{recoverbyaverage}.) Thus Leegwater proves that Parameter Independence and `no-conspiracy', taken together, are enough to imply that the hidden variables must be otiose or trivial in the sense above (2016: Theorem 1, p. 21). This is a stunning no-go result.\footnote{Furthermore, Leegwater  allows the measure over hidden variables to depend on more than just the quantum state (although not which quantity is measured). For example, it can depend on the specific method that was used to prepare the state. 

I stress that discerning the no-conspiracy assumption is  not original to me: Landsman and Leegwater are perfectly clear that they make this assumption, even though it is not given a formal label or acronym  (Landsman 2015, p. 122103-2, assumption CQ and its footnote 14; 2017, p. 221, Definition 6.20 and following text; Leegwater, 2016, p. 21 footnote 7 and its preceding text).}  

Thus, putting the two points together: while Colbeck and Renner gave a dazzlingly inventive but heuristic derivation, the papers by Landsman and  Leegwater  are also dazzling examples---of mathematical invention, as well as rigour. \\

What is the upshot as regards Kent? At first sight, it seems that his proposal {\em does} satisfy no-conspiracy. For he averages over the possible values of his hidden variables, i.e. over the possible final boundary conditions, with the Born-rule probabilities given by the final universal quantum state  $| \psi_S \rangle$. And $| \psi_S \rangle$ is independent of which quantity, or quantities, were previously measured, even though it is the universal state. Agreed: there is a fact about which quantity got measured (and about what the outcome was), according to Kent's recovery of a quasiclassical world (Sections \ref{recover} and \ref{2constraints}). But these facts leave no `mark' on---have no back-reaction on---the universal quantum state, which always evolves unitarily. Thus in the Schr\"{o}dinger picture, $| \psi_S \rangle$ gets, at an intermediate time when an experiment is set up, components (non-zero amplitudes) for various possible choices of quantity (settings) and, soon thereafter, components  for various possible outcomes. (Think, if you will, of an Everettian's description of the setting-up, and performance, of a run of an experiment: the components correspond to the Everettian's branches.) So it seems that Kent's proposal satisfies no-conspiracy; and so, thanks to Leegwater's theorem, must violate Parameter Independence at the micro-level---giving a clear analogy  with the pilot-wave theory.\footnote{Note the obvious contrast with the dynamical reduction programme, i.e. with the view that the universal quantum state evolves {\em non}-unitarily, collapsing appropriately throughout history so as to yield a quasiclassical world (cf. footnote 1 and Section \ref{compel}). On that view, the quasiclassical world no doubt includes which quantity, or quantities, are measured at all the various times, and so the final universal quantum state certainly {\em will} depend on such facts. So in such a universe, a Kentian algorithm applied to the final condition  would violate no-conspiracy.}

But on reflection, it is {\em not} clear that Kent's proposal obeys no-conspiracy in the relevant sense, i.e. for the measure used in deriving a Bell inequality (cf. Section \ref{PIOI}: here, I am indebted to Gijs Leegwater). For {\em a priori}, this measure need not be the Born-rule probabilities given by the final universal quantum state $| \psi_S \rangle$. In particular: since (we can presume) a Kentian toy model of a Bell experiment  makes the final condition ``rich enough'' to determine which quantity was measured earlier, the relevant measures for different settings may have disjoint supports. In short: about Parameter Independence, the jury is still out.

\subsection{Mandatory peaceful coexistence? A peace-pipe for Shimony}\label{compel}
I end with one last, happy, point: in effect, a peace-pipe to smoke with my dedicatee, Abner Shimony. Namely: these theorems, especially Leegwater's theorem, promote peaceful coexistence of the kind Shimony sought---while leaving us plenty of work for the future, even if we join Shimony in favouring some process of dynamical reduction. 

For recall that their gist is that Parameter Independence is enough to ban hidden variable supplementations. Thus Shimony's overall view of non-locality---accepting Outcome Dependence so as to avoid a Bell inequality, and seeing no reason {\em not} to endorse Parameter Independence---turns out, by dint of the hard work of these theorems, to lead to `unsupplemented' quantum theory.

This is an important strengthening of the argument for peaceful coexistence. For the traditional Bell-Shimony argument is, essentially, that Parameter Independence and empirical adequacy for the Bell experiment imply Outcome Dependence: and we then worry about whether Outcome Dependence really satisfies the spirit of relativity (Bell himself thinking `No'; cf. Section \ref{preamble}). But these new theorems imply that Parameter Independence and empirical adequacy (admittedly: in other nonlocality experiments---but ones where we are confident of quantum theory) imply `unsupplemented' quantum theory: a much stronger conclusion that Outcome Dependence---which, in a relativistic world, amounts to mandatory peaceful coexistence.  

Of course, this conclusion is not the last word. This new peaceful coexistence may be mandatory, given just Parameter Independence. But one naturally wants it to be spelt out in more detail: a project which, I argued in Section \ref{preamble2.3}, eventually requires a solution, or at least a sketched solution, of the measurement problem. 

To honour Shimony's memory, I shall end by briefly discussing this in relation to his preferred approach: dynamical reduction (i.e. non-unitary evolution for strictly isolated systems; cf. the references in footnote \ref{DRP}), rather than postulating  extra values for quantities, as in the pilot-wave theory.

At first sight, this conclusion  sits well with Shimony's preference. After all, dynamical reduction models do not try to recover the Born-rule probabilities of outcomes by probabilistically averaging over a hidden variable. But there is a subtlety, indeed work to be done, hereabouts. For one {\em can} think of such a model  as recovering probabilities of outcomes by probabilistic averaging over realizations of the stochastic noise, i.e. over  how the stochastic noise `happens to go'. For recall that once given:\\
\indent (i) an initial quantum state $\psi$, and \\
\indent (ii) a choice of measured quantity $Q$, and more specifically \\
\indent (iii) an  apparatus in a specified state, and a total Hamiltonian for the joint system comprising micro-system and apparatus; then\\
each realization of the noise leads to one definite measurement outcome.\\
So thinking of each realization as a hidden variable $\lambda$, these hidden variables are deterministic; so they are certainly not trivial in these theorems' sense.   But  the measure used in the probabilistic averaging over realizations is judiciously (brilliantly!) defined so as to be sensitive to $\psi$'s Born-rule probabilities for values of $Q$, in just such a way as to recover the Born-rule probabilities for outcomes of the measurement process. Thus the measure depends on which quantity is chosen to be measured. So the  point here is that the theorems of Colbeck, Renner and Leegwater are evaded in the sense that their `no-conspiracy' assumption, that the measure over hidden variables be independent of the quantity measured, is violated. 

But again, this is not the last word. Clearly, there is work to be done here, relating the notions of one's favoured dynamical reduction model to the notions and assumptions of the theorems. In particular: suppose one considers instead the measures over the noise that are independent of the quantity measured, and even independent of the quantum state---which are, after all, explicitly there in the formalism of dynamical reduction models. Yet it still seems reasonable to think of the model as underpinning orthodox quantum probabilities, with each realization of the noise being a (deterministic) hidden variable. So one faces the question: what assumptions of the theorems of Colbeck, Renner and Leegwater are now violated? Perhaps Parameter Independence? Work for the future!

\section{Summary}\label{summary}
My topic has been whether there can be `peaceful coexistence' between quantum theory and special relativity; and in particular, Shimony's hope that  Outcome Dependence would show a way---maybe the best or only way---to secure it.

In Section \ref{preamble2}, I rehearsed the issues, and concluded sceptically that accepting Outcome Dependence is far from enough to get peaceful coexistence. In short, one needs a relativistic solution to the quantum measurement problem: given such a solution, one can then assess whether Outcome Independence fails---and agreed, one would expect it to fail. 

Then I turned to one proposed solution: i.e. one realist one-world Lorentz-invariant interpretation of quantum theory---namely Kent's recent proposal (2014, 2015, 2016).  I first  reported his main ideas (Section \ref{intro}). Then in Section \ref{od}, I spelt out how, with his proposed beable making precise what is an outcome of a measurement, his interpretation is like the pilot-wave theory, in that it satisfies Outcome Independence at the micro-level, and (presumably) Outcome Dependence at the observable level. 

Then in Section \ref{pi?}, I reported recent remarkable theorems by Colbeck, Renner and Leegwater which bear on whether his interpretation obeys Parameter Independence at the micro-level. And I argued that these theorems make some kind of peaceful coexistence mandatory for someone who, like Shimony, endorses Parameter Independence.  

So the upshot is that Shimony's hope for peaceful coexistence is alive and well. But to come true, it will need more than a judicious  choice of which assumption of a Bell theorem to deny. It will probably need  an agreed relativistic solution to the quantum measurement problem. In seeking such agreement, we need both to assess proposed solutions like Kent's, and to analyse the constraints on the solution  implied by general theorems like those of Colbeck, Renner and Leegwater. \\ \\ \\

\noindent {\em Acknowledgements}:--- Dedicated to the memory of Abner Shimony. As all who met him soon realized, it was a pleasure and a privilege to know him, both as a person and as an intellect. It is a pleasure to thank Masanao Ozawa for the invitation to the Nagoya conference; and to thank him, Francesco Buscemi and the other local organizers, for a very enjoyable and valuable meeting. I am very grateful to Adrian Kent for generous advice and encouragement; and to Bryan Roberts for the splendid diagram. For comments on a previous version, I thank an anonymous referee, audiences in Cambridge, Wayne Myrvold, James Read, Bryan Roberts and four Dutch wizards: Dennis Dieks, Fred Muller and especially Guido Bacciagaluppi and Gijs Leegwater.

\newpage

\section{References}

\indent Allahverdyan, A.,  Ballian R. and  Nieuwenhuizen T.: Understanding quantum measurement
from the solution of dynamical models. {\em Physics Reports} {\bf 525} 1-166  (2013).


Barrett, J.: {\em The Quantum Mechanics of Minds and Worlds}, Oxford University Press; (1999).

Bassi, A. and  Ghirardi, G.C.: Dynamical reduction models. {\em Physics Reports} {\bf 379}, 257-426 (2003).

Bell, J: {\em Speakable and Unspeakable in Quantum Mechanics}, Cambridge: University Press; second edition; (2004). 

Bell, J. and Shimony, A., Horne, M. and Clauser, J.: An exchange on local beables, {\em Dialectica} {\bf 39}, 85-1110  (1985).

Berndl, K: Global existence and uniqueness of Bohmian trajectories, In Cushing J.,  Fine A. and Goldstein S. (eds), {\em Bohmian Mechanics and Quantum Theory: An Appraisal},  Kluwer Academic; arxiv: quant-ph/9509009 (1996).

Berndl, K., D\"{u}rr D., Goldstein S., Peruzzi G. and Zanghi, N: On the global existence of Bohmian mechanics, {\em Communications in Mathematical Physics} {\bf 173}, 647-673  (1995).

Bohm, D. and Hiley, B. {\em The Undivided Universe}, London: Routledge (1992).

Bricmont, J.:  {\em Making Sense of Quantum Mechanics}, Springer (2016).

Brown, H. and Timpson, C.: Bell on Bell's theorem: the changing face of nonlocality, in M. Bell and Shan Gao (eds), {\em Quantum Nonlocality and Reality: Fifty Years of Bell's Theorem}, Cambridge University Press; arxiv: 1501.03521: (2016).

Brown, H. and Wallace D.: Solving the measurement problem: de Broglie-Bohm loses out to Everett,  {\em Foundations of Physics}  {\bf 35}, 517-540 (2005). 

Butterfield, J .: Bell's Theorem: what it takes, {\em British Journal for the Philosophy of Science}, {\bf 43}, 41-83 (1992).

Butterfield, J .:  Stochastic Einstein Locality Revisited, {\em British Journal for the Philosophy of Science}, {\bf 58}, 805-867 (2007).

Butterfield, J .:  Assessing the Montevideo interpretation of quantum mechanics, {\em Studies in the History and Philosophy of Modern Physics}, {\bf 52A}, 75-85; \\
At: http://arxiv.org/abs/1406.4351; http://philsci-archive.pitt.edu/10761/; (2015).

d'Espagnat. B.:  {\em Conceptual Foundations of Quantum Mechanics} Reading, Mass: Benjamin; second edition (1976).

Clauser, J. and Horne, M.: Experimental consequences of objective local theories, {\em Physical Review D} {\bf 10} 526-534 (1974). 

Clifton, R. and Jones, M.: Against experimental metaphysics. In: French, P., Euling, T. and Wettstein, H. (eds.) {\em Midwest Studies in Philosophy}, volume 18: Philosophy of Science; Minneapolis: University of Minnesota Press; 295-316 (1993).

Colbeck, R. and Renner, R.: No extension of quantum theory can have improved predictive power, {\em  Nature Communications} {\bf 2}, 411. http://dx.doi.org/10.1038/ncomms1416 (2011).

Colbeck, R. and Renner, R.: The completeness of quantum theory for predicting measurement outcomes.  arxiv:1208.4123 (2012).

 Cushing, J. and McMullin, E. (eds): {\em Philosophical Lessons from Quantum Theory}, University of Notre Dame Press (1989).
 
Dewdney, C., Holland P. and Kyprianidis, A.: A causal account of non-local Einstein-Podolsky-Rosen spin correlations, {\em Journal of Physics A: Math. Gen.} {\bf 20}, 4717-4732 (1987).

Dieks, D and Vermaas, P. (eds.): {\em The Modal Interpretation of Quantum Mechanics}, Kluwer (1998).

Friederich, S.: {\em Interpreting Quantum Theory: a Therapeutic Approach}, Palgrave Macmillan (2015).

Ghirardi, G.: The interpretation of quantum mechanics: where do we stand?, {\em Journal of Physics: Conference Series} {\bf 174}  (DICE 2008) 012013 (2009).

Ghirardi, G., Rimini, A. and Weber, T:  A general argument against superluminal transmission through the quantum mechanical measurement process, {\em Lettere al Nuovo Cimento} {\bf 27}, 293-298 (1980).

Healey, R: Quantum theory: a pragmatist approach, {\em British Journal for the Philosophy of Science}  {\bf 63}, 729Ð771 (2012).

Healey, R: How quantum theory helps us explain, {\em British Journal for the Philosophy of Science}  {\bf 64}, 1-43; http://dx.doi.org/10.1093/bjps/axt031; (2013).

Healey, R: Causality and chance in relativistic quantum field theories, {\em Studies in the History and Philosophy of Modern Physics} {\bf 48}, 156-167 (2014).

Healey, R.:  {\em The Quantum Revolution in Philosophy}, Oxford University Press; (2017).

Henson, J.: Non-separability does not relieve the problem of Bell's theorem, {\em Foundations of Physics} {\bf 43}, 1008-1038 (2013). 

Holland, P.: {\em The Quantum Theory of Motion},  Cambridge: University Press; (1993). 

Howard, D.: Holism, separability and the metaphysical implications of the Bell experiments. In: Cushing  and McMullin (eds.) (1989), pp. 224-253; (1989). 

Isham, C: {\em Lectures on Quantum Theory}, London; Imperial College Press (1995).

Janssen, H.: Reconstructing Reality: Environment-Induced Decoherence, the Measurement Problem, and the Emergence of Definiteness in Quantum Mechanics. http://philsci-archive.pitt.edu/4224/ ; (2008).

Jarrett, J.: On the physical significance of the locality conditions in the Bell arguments, {\em Nous} {\bf 18}, 569-589 (1984).

Jarrett, J.: On the separability of physical systems, in Myrvold, W. and Christian, J. (eds.) (2009); pp. 105-124: (2009).

Jordan, T.: Quantum correlations do not transmit signals, {\em Physics Letters} {\bf 94A}, 264 (1983).

Kent, A.: Solution to the Lorentzian quantum reality problem, {\em Physical Review A} {\bf 90}, 012107; arxiv: 1311.0249; (2014). 

Kent, A.: Lorentzian quantum reality: postulates and toy models, {\em Philosophical Transactions of the Royal Society A} {\bf 373}, 20140241; arxiv:  1411.2957; (2015).

Kent, A.: Quantum reality via late time photodetection; arxiv:  1608.04805 (2016).

Landsman, N.: Spontaneous symmetry breaking in quantum systems: emergence or reduction?
{\em Studies in History and Philosophy of Modern Physics} {\bf 44(4)}, 379-394 (2013).

Landsman, N. and Reuvers, R:  A flea on Schr{\"{o}}dinger's cat. {\em Foundations of Physics} {\bf 43},
373-407; (2013a).

Landsman, N.: The Colbeck-Renner theorem, {\em Journal of Mathematical Physics} {\bf 56},
122103; (2015). 

Landsman, N.: {\em Foundations of Quantum Theory}, Springer (2017). Open access: available at: https://link.springer.com/book/10.1007/978-3-319-51777-3

Leegwater, G.: An impossibility theorem for parameter independent hidden variable theories, {\em Studies in the History and Philosophy of Modern Physics}, {\bf 54} 18-34; http://philsci-archive.pitt.edu/12067/;  (2016).

Leggett, A.: Probing Quantum Mechanics Towards the Everyday World: Where do we Stand?, {\em Physica Scripta}  {\bf T102}, 69-73, (2002).

Morganti, M. A new look at relational holism in quantum mechanics, {\em  Philosophy
of Science} {\bf 76}, 1027Ð1038 (2009).

Muller, F.: The locality scandal of quantum mechanics. In: Dalla Chiara, M. et al. (eds.), {\em Language, Quantum, Music}, Synthese Library volume 281, Dordrecht: Kluwer Academic, 241-248 (1999).

Myrvold, W.: On peaceful coexistence: is the collapse postulate incompatible with relativity? {\em Studies in the History and Philosophy of Modern Physics} {\bf 33}, 435Ð466  (2002).

Myrvold, W.:  Relativistic quantum becoming, {\em British Journal for the Philosophy of Science} {\bf 54}, 475Ð500 (2003).

Myrvold, W.: Lessons of Bell's theorem: Nonlocality, yes; Action at a distance, not necessarily, in M. Bell and Shan Gao (eds), {\em Quantum Nonlocality and Reality: Fifty Years of Bell's Theorem}, Cambridge University Press. Available at: http://philsci-archive.pitt.edu/12382/ ; (2016).

Myrvold, W.: Ontology for collapse theories, forthcoming in  Shan Gao (ed.) {\em Collapse of the Wave Function}, Cambridge University Press (2017).

Myrvold, W. and Christian, J. (eds.): {\em Quantum Reality, Relativistic Causality, and Closing the Epistemic Circle: Essays in Honor of Abner Shimony}, Springer (2009).

Norsen, T: Local causality and completeness: Bell vs. Jarrett, {\em Foundations of Physics} {\bf 39}, 273-294 (2009).

Norsen, T: John S. Bell's concept of local causality, {\em American Journal of Physics} {\bf 79}, 1261-1275 (2011). 

Pearle, P.: How stands collapse II, in Myrvold, W. and Christian, J. (eds.) (2009); pp. 257-292: (2009).

Redhead, M: {\em Incompleteness, Nonlocality and Realism} Oxford: University Press (1987). 


Schr\"{o}dinger, E.: The Present Situation in Quantum Mechanics: A Translation of Schršdinger's "Cat Paradox" Paper 
(trans: J D. Trimmer) {\em Proceedings of the American Philosophical Society,} {\bf 124}, (Oct. 10, 1980), pp. 323-338;  
American Philosophical Society; http://www.jstor.org/stable/986572; (1935).

Shimony, A.: Controllable and uncontrollable nonlocality. In: Kamefuchi , S. et al. (eds) {\em Foundations of Quantum Mechanics in the Light of New Technology}, Tokyo: Physical Society of Japan; reprinted in Shimony (1993), 130-139: page references to reprint; (1984). 

Shimony, A.: Events and processes in the quantum world. In:  Penrose, R. and Isham C. (eds.) {\em Quantum Concepts in Space and Time} Oxford: University Press; reprinted in Shimony (1993), 140-162: page references to reprint; (1986).

Shimony, A.: Desiderata for a modified quantum dynamics. In:   (eds.) {\em PSA 1990} volume 2; Proceedings of the 1990 meeting of the Philosophy of Science Association; East Lansing, Michigan: Philosophy of Science Association; reprinted in Shimony (1993), 55-67: (1990).

Shimony, A.: {\em Search for a Naturalistic World View: Volume II: natural science and metaphysics}. Cambridge: University Press (1993).

Shimony, A.: Unfinished work, a bequest. In: Myrvold and Christian (eds.) (2009); pp. 479-491 (2009).

Shimony, A.: Bell's theorem, in {\em The Stanford Encyclopedia of Philosophy}. Available at: https://plato.stanford.edu/entries/bell-theorem/ . (2009a).

Teller, P.: Relativity, relational holism and the Bell inequalities. In: Cushing and McMullin (eds.) (1989); pp. 208-223 (1989). 

Valentini, A.: Signal locality in hidden variable theories, {\em Physics Letters A} {\bf 297}, 273-278; arxiv: quant-ph/0106098; (2002).

Valentini, A.: Signal locality and sub-quantum information in deterministic hidden variable theories, in T. Placek and J. Butterfield (eds.)  {\em Non-Locality and Modality} NATO Science Series II: volume 64, Kluwer; arxiv: quant-ph/0112151; (2002a).

van Fraassen, B.: The charybdis of realism: epistemological implications of Bell's theorem, {\em Synthese} {\bf 52}; 25-38 (1982).

von Neumann, J.: {\em Mathematical Foundations of Quantum Mechanics}, Princeton: University Press (English translation 1955, reprinted in the Princeton Landmarks series 1996); (1932). 

Wallace, D.:  {\em The Emergent Multiverse}, Oxford University Press; (2012).

Wiseman, H. and Cavalcanti, E.: {\em Causarum Investigatio} and the two Bell's theorems of John Bell, in {\em Quantum Unspeakables II}, ed. R. Bertlmann and A. Zeilinger, Springer; pp. 119-142; arxiv: 1503.06413; (2017).

Zeh, H-D, Joos, E. et al.; {\em Decoherence and the Appearance of a Classical World in Quantum Theory},  second edition; Springer (2003).
 
\end{document}